\documentclass[twocolumn]{aastex631}
\usepackage[]{amsmath,graphicx,natbib,url,multirow,float}
\submitjournal{Astrophysical Journal}

\begin{document}
\newcommand{\kms}{\hbox{km\,s$^{-1}$}}
\newcommand{\Msun}{\mbox{$M_{\sun}$}}
\newcommand{\Lsun}{\mbox{$L_{\sun}$}}
\newcommand{\cm}[1]{\mbox{cm$^{#1}$}}
\newcommand{\cc}{\mbox{cm$^{-3}$}}
\newcommand{\um}{\mbox{$\mu$m}}

\shorttitle{Impact of Thermal and Abundance Profile Assumptions}
\shortauthors{Rowland et al.}

\title{Towards Robust Atmospheric Retrieval on Cloudy L Dwarfs: \\The Impact of Thermal and Abundance Profile Assumptions}

\author[0000-0003-4225-6314]{Melanie J. Rowland}
\affil{University of Texas at Austin, Department of Astronomy, 2515 Speedway C1400, Austin, TX 78712, USA}

\author[0000-0002-4404-0456]{Caroline V. Morley}
\affil{University of Texas at Austin, Department of Astronomy, 2515 Speedway C1400, Austin, TX 78712, USA}

\author[0000-0002-2338-476X]{Michael R. Line}
\affil{School of Earth $\&$ Space Exploration, Arizona State University, Tempe AZ 85287, USA}

\correspondingauthor{Melanie Rowland}
\email{mrowland@utexas.edu}

\begin{abstract}
Constraining L dwarf properties from their spectra is challenging. Near-infrared spectra probe a limited range of pressures, while many species condense within their photospheres. Condensation creates two complexities: gas-phase species ``rain out'' (decreasing in abundances by many orders of magnitude) and clouds form. We designed tests using synthetic data to determine the best approach for retrieving L dwarf spectra, isolating the challenges in the absence of cloud opacity. We conducted atmospheric retrievals on synthetic cloud-free L dwarf spectra derived from the Sonora Bobcat models at SpeX resolution using a variety of thermal and chemical abundance profile parameterizations. 
For objects hotter than L5 ($T_{\rm eff}\sim 1700$ K), the limited pressure layers probed in the near-IR are mostly convective; parameterized PT profiles bias results and free, unsmoothed profiles should be used. Only when many layers both above and below the radiative-convective boundary are probed can parameterized profiles provide accurate results. 
Furthermore, a nonuniform abundance profile for iron hydride (FeH) is needed to accurately retrieve bulk properties of early- to mid- L dwarfs. Nonuniform prescriptions for other gases in near-IR retrievals may also be warranted near the L/T transition (CH$_{4}$) and early Y dwarfs (Na and K). We demonstrate the utility of using realistic self-consistent models to benchmark retrievals and suggest how they can be used in the future. 
\end{abstract}

\section{Introduction}\label{introduction} 
Atmospheric studies of brown dwarfs and exoplanets are used to constrain key atmospheric properties that can reveal information about an object's history. Ages of brown dwarfs can be inferred from effective temperatures and surface gravities \citep{burrows2001, saumonmarley2008}, while metallicities, C/O ratios, and recently carbon isotope ratios have been used to constrain giant planet formation pathways, formation location, and migration \citep{oberg2011, Madhusudhan2014ApJ...794L..12M, Madhusudhan2017MNRAS.469.4102M, zhangy2021_isotopejupiter}. Brown dwarfs offer the opportunity to study atmospheres with similar chemical complexity to giant planets while removing complications from stellar irradiation and prohibitively low signal to noise (SNR). 

Inferring bulk properties of brown dwarfs and exoplanets has been done by comparing the object's near-infrared (NIR) spectrum to model spectra generated from theoretical models \citep{cushing2008}. One of the two prevailing modeling techniques is self-consistent models computed on grids of effective temperature and surface gravity. Early brown dwarf models assumed radiative-convective thermochemical equilibrium and were either adapted from solar system models for hotter and more massive objects \citep{marley1996} or from cool star models for cooler and less massive objects \citep{burrows1993}. These models have evolved in complexity to include clouds \citep{ackermanmarley2001}, rainout of gases \citep{loddersfegley2002, marley2021ApJ...920...85M}, and disequilibrium chemistry \citep{phillips2020, mukherjee2022, lacy2022}, and rely on assumed atmospheric chemistry paradigms and 1D radiative-convective equilibrium \citep{burrows2001,marleyrobinson2015} to reduce the dimensionality of the models.

The inclusion of more complex atmospheric processes in grid models has resulted in improved data-model agreement and granted key insights into the chemical and physical phenomena at play. However, the nature of grid modeling limits inferences regarding object properties to only those specified as grid dimensions. Remaining data-model mismatches across all existing grid models imply that some assumptions required by grid modeling may not be valid in many objects.

Atmospheric retrievals provide better model-data fits and determination of atmospheric properties not included in grid models. The better model fits come from relaxing some of the assumptions that are needed in self-consistent models and replacing them with many parameters that are directly determined. These better data-model fits come with several risks. The larger number of parameters can lead to over-fitting, and the relaxation of parameters can lead to unphysical results, so retrieval parameterization and interpretation require careful consideration. Two important parameterization choices include the thermal profile and the chemistry. 

\subsection{Thermal Profile Parameterization in Retrievals}
Thermal profile parameterizations must allow some degree of flexibility while producing physically realistic profiles. Less complex and less flexible profiles make more assumptions about the structure of the atmosphere, but require fewer parameters, which may be warranted for low SNR or low resolution observations. Profiles used in retrievals include a 5-parameter joint exponential power law \citep{madhusudhanseager2009,burningham2017}, 6-parameter radiative-convective power law with flexible upper atmospheres \citep{gravity2020,molliere2020}, variable parameter (2 - $\infty$) piece-wise polynomial profile \citep{kitzmann2020heliosr2}, and 17-parameter free profile \citep{line2015}. 

\subsection{Chemistry Parameterization in Retrievals}
    Similarly, the chemistry parameterization has fallen into two categories: equilibrium chemistry and ``free'' chemistry. Most equilibrium chemistry retrievals typically only retrieve the bulk metallicity and the C/O ratio and then use the thermal profile to obtain abundance profiles by either interpolating from a precomputed equilibrium chemistry grid or calculating them directly using an equilibrium chemistry code like \texttt{FastChem} \citep{stock2018fastchem}. The later method allows changing element abundances instead of assuming solar ratios. This type of parameterization is useful when only a few model parameters are justified due to  sparse or low SNR spectra, such as in \cite{molliere2020}, but its ability to accurately characterize atmospheres relies heavily on model assumptions, as in grid models. In general, hot atmospheres tend to relax into a state of chemical equilibrium because the chemical timescale is short at high temperatures. Departures from chemical equilibrium can be incorporated with quench pressure approximations and rainout condensation can be incorporated into equilibrium chemistry grids,  but conclusions about chemical abundances and gas ratios different from solar ratios are still limited. Alternatively, the free chemistry approach directly retrieves abundances for a determined subset of gases, in which the abundance profile of each trace species is retrieved independently by one constant-with-altitude (or ``uniform'') parameter per species. \citet{changeat2019} and \citet{Bourrier2020} employed nonuniform approaches, but required 5 or 4 parameters for a nonuniform gas, respectively. The flexibility of free chemistry approaches allows retrievals to characterize more molecular and atomic ratios, while the drawbacks include a large number of parameters, retrieving non-physical abundances, and not capturing nonuniform-with-pressure abundance profiles that may be important when rainout chemistry impacts prominent opacity sources.

\subsection{Prior Brown Dwarf Retrieval Results}
The increased flexibility provided by these thermal and chemistry profile parameterizations have allowed retrievals to identify regimes where grid model assumptions break down and to identify missing model physics. Free chemistry retrievals allow carbon, nitrogen, and oxygen reservoir gases as well as alkalis to vary independently, in contrast to grid models that require abundances change in concert to maintain chemical equilibrium. \citet{line2017} and \citet{zalesky2019} showed that retrieved abundances in mid-T dwarfs can largely be explained by chemical equilibrium, but retrievals in the late-L dwarf regime by \citet{burningham2017} indicate that either chemical disequilibrium via vertical mixing or nonsolar CNO ratios are needed to explain retrieved abundances. \citet{line2017} and \citet{zalesky2019} also showed decreasing alkali abundances starting in objects with T$_{\rm eff}$ = 1000 K through the late-T and early-Y dwarf sequence. This indicated a departure from pure local thermo-chemical equilibrium models which predicted decreasing alkali abundances by 1200 K, and instead favored rainout chemistry as predicted by \citet{lodders1999} and \citet{burrows2001}.

\subsection{The Challenges of L Dwarf Retrievals}
Flexible thermal profile retrievals have shown that radiative-convective equilibrium is valid in T and Y dwarfs \citep{line2015, line2017, kitzmann2020heliosr2,zalesky2022}, however many retrieval studies of L dwarfs have favored isothermal temperature profiles \citep{burningham2017, lueber2022, gonzales2022arXiv220902754G}. Multiple explanations for this preferred profile have been suggested. One explanation is the presence of clouds which add grey opacity at the observed NIR wavelengths. These clouds mimic an isothermal profile and obscure flux from deeper layers. \citet{tremblin2016} suggests an alternative to the L dwarf cloud paradigm to explain the isothermal profiles by invoking an adiabatic index lowered by thermo-chemical convection arising from the transition from CO/CH$_{4}$ in an out-of-equilibrium atmosphere. 

These open questions surrounding L dwarf properties highlight the difficulty in expanding retrieval frameworks to hotter temperatures. While these thermal and abundance profile parameterizations have been successful at retrieving cloud-free late-T benchmarks like Gl-570D \citep{line2015,burningham2017,kitzmann2020heliosr2} and retrieving most bulk properties up the T dwarf sequence \citep{line2017,zalesky2022, lueber2022} and down the Y dwarf sequence \citep{zalesky2019}, the validity of these parameterizations and their effects on retrieved bulk properties have not been robustly tested at hotter temperatures. 

Three main challenges exist in NIR modeling of L dwarfs: 1.) NIR opacity windows that probe many layers of the atmosphere in cooler objects close with the rise of hydride and oxide opacities. NIR spectra of L dwarfs probe only a narrow range of the atmosphere, which inhibits robust thermal profile characterization.  2.) Prominent opacity sources at cooler temperatures are expected to have uniform-with-pressure abundance profiles, while prominent absorbers at hotter temperatures like hydrides and oxides are heavily impacted by rainout, making their abundance profiles strongly nonuniform. 3) The presence of clouds increasingly impacts mid- to late- L dwarf spectra and complicates analyses because they are degenerate with an isothermal profile and obscure atmospheric structure below the cloud layers. 

In light of these challenges, we aim to assess whether our existing thermal and chemical profile parameterizations are adequate in retrieving accurate bulk properties. In this paper, we tackle two of the three challenges. First, which PT profile parameterizations accurately retrieve bulk properties and thermal profiles with spectral information from limited pressure ranges? Second, what kind of free chemistry profile parameterizations accurately retrieve bulk properties in objects with abundances that are strongly nonuniform? 

To do this, we perform a suite of retrievals using a variety of thermal and abundance parameterizations on synthetic benchmarks generated from self-consistent grid model spectra for which we know ground-truth properties. While real mid- to late- L dwarfs are expected to have clouds, we retrieve on cloud-free synthetic spectra to isolate the effects the other two challenges pose to our retrieval parameterizations. In Section \ref{data}, we detail how we generate the synthetic benchmarks, and in Section \ref{retrievalframework} we detail the retrieval framework and the tested thermal and chemistry parameterizations. In Section \ref{Results} we show results and the winning parameterizations. Finally, in Section \ref{Discussion} we discuss how this informs the validity of model assumptions in different regimes. 

\section{Methods}\label{Methods}
Retrieving on synthetic spectra we generate ourselves allows us to know ground truth values of not only bulk properties like effective temperature, surface gravity, metallicity, and C/O ratios, but also the true thermal profile and abundance profiles, something that isn't possible with real benchmark objects. We describe our synthetic spectra in Section \ref{data} and the suite of retrievals with varying parameterizations we performed on them in Section \ref{retrievalframework}.

\subsection{Data}\label{data}
\begin{figure*}[!ht]
    \centering
    \includegraphics[width=1.0\textwidth]{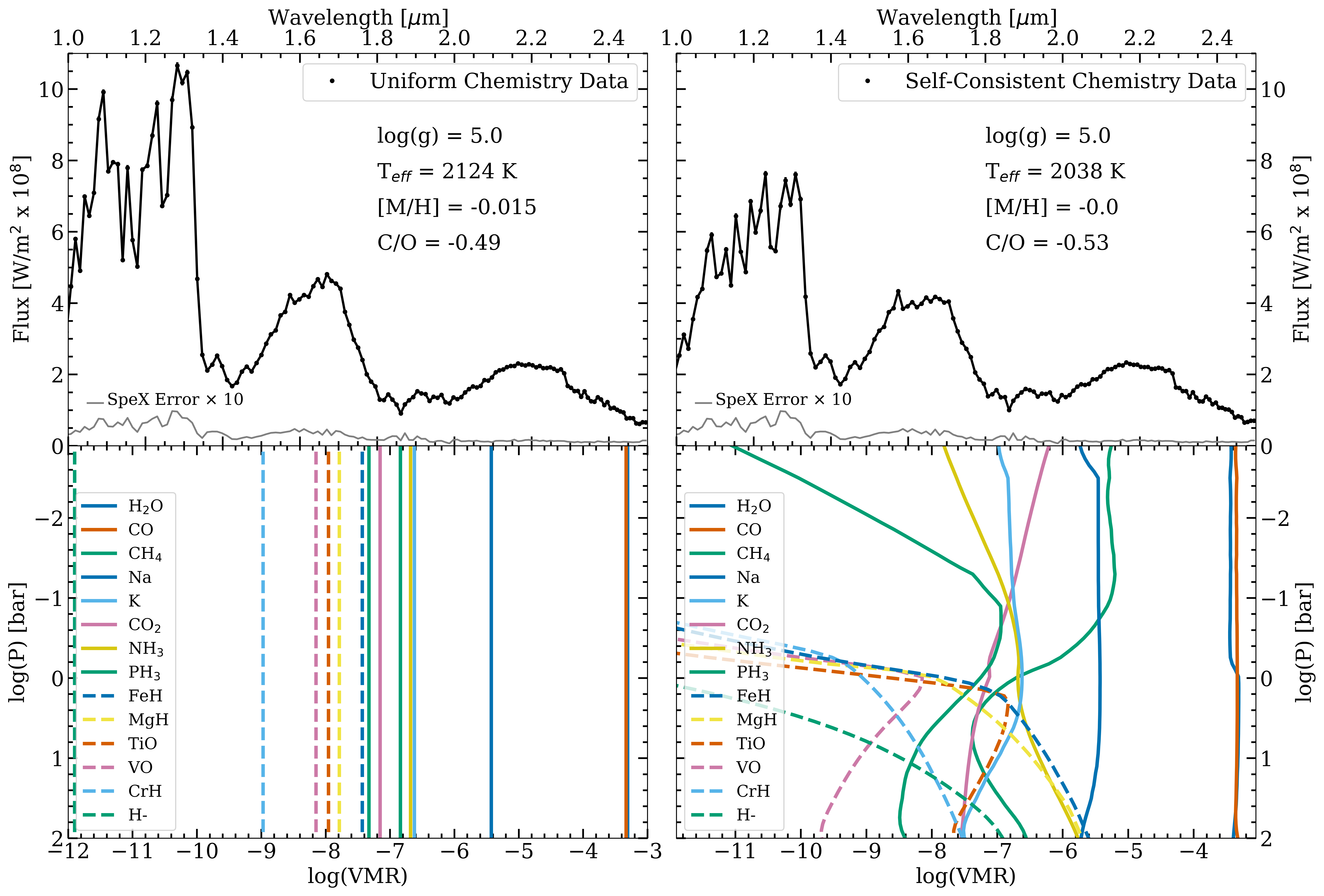}
    \caption{L2 (T$_{\rm eff}$ = 2000 K) synthetic spectra [top] generated with uniform chemistry [left] or self-consistent chemistry [right] in black with characteristic SpeX error $\times$ 10 in grey for visibility. The chemistry profiles for the 14 gas species used to generate each spectrum shown on bottom.}
    \label{fig:uniformvssc}
\end{figure*}
The CHIMERA retrieval framework’s forward model was used to create two synthetic $\sim$L2 spectra based on the self-consistent cloud-free Sonora Bobcat model's T$_{\rm eff}$ = 2000 K, log(g) = 5.0, solar metallicity, solar C/O object \citep{marley2021ApJ...920...85M} at 10 parsecs. Both spectra use the Sonora model's PT profile (91 layers, log(P) = -3.75 to 1.643 bars) and 14 selected gas species. The gas species included are H$_{2}$O, CO, CO$_{2}$, CH$_{4}$, NH$_{3}$, Na, K, PH$_{3}$, TiO, VO, FeH, CrH, MgH, and H-, along with H$_{2}$-H$_{2}$ and H$_{2}$-He collision induced absorption opacities. We use the absorption cross section sources from the \citet{marley2021ApJ...920...85M} Sonora Bobcat models for these 14 gases and provide the molecular cross section sources in Table \ref{tab:opacities}.
\begin{deluxetable}{cc}
\tablecolumns{2}
\tablecaption{Molecular Opacity Sources\label{tab:opacities}}
\tablewidth{0pt}
\tablehead{
\colhead{Species} & \colhead{Source}}
\startdata
H$_{2}$O & \cite{Tennyson2018}; \cite{Barber2006}\\
CH$_{4}$ & \cite{Yurchenko2013}; \cite{Yurchenko2014}\\
CO & H10$^{a}$; \cite{li2015} \\
CO$_{2}$ & \cite{Huang2014} \\
NH$_{3}$ & \cite{Yurchenko2011}\\
PH$_{3}$ & \cite{Sousa-Silva2015}$^{b}$\\
TiO & \cite{Schwenke1998}; \cite{Allard2000} \\
VO &  \cite{mckemmish2016}; ExoMol$^{b}$\\
MgH & \cite{weck2003}$^{c}$\\
CrH &  \cite{burrows2002}\\
FeH & \cite{dulick2003}; \cite{hargreaves2010}\\
\enddata
\tablecomments{$^{a}$HT10 = HITEMP 2010 \citep{rothman2010}: http://www.cfa.harvard.edu/hitran/HITEMP.html; $^{b}$http://www.exomol.com (\cite{Tennyson2012}).; $^{c}$http://www.physast.uga.edu/ugamop/.}
\end{deluxetable}

The first L2 spectrum was designed to isolate the effects of narrow pressure ranges probed due to increased hydride and oxide opacity at high temperatures. This spectrum was generated with uniform-with-pressure chemistry profiles for the 14 gas species. Self-consistent chemistry profiles from the Sonora models were analyzed and mixing ratios for each gas were held constant at their value at 1 bar, corresponding to the values in the middle of the photosphere. This uniform chemistry spectrum was used to test different PT profile parameterizations and to inform the parameterizations of subsequent retrievals. In addition to this L2, T$_{\rm eff}$ = 2000 K spectrum, we also generated two additional uniform chemistry spectra (L5, T$_{\rm eff}$ = 1700 K and L7, T$_{\rm eff}$ = 1600 K) using the same approach to explore the effects of PT profile parameterizations across a multiple spectral types. 

The second L2 spectrum was designed to include the effects of nonuniform chemistry profiles, which are expected to occur with rainout chemistry. Mixing ratios for each gas were determined by the self-consistent chemistry profiles from the Sonora model. This self-consistent chemistry spectrum was used to test different free chemistry parameterizations. 

The output spectra were binned to IRTF SpeX (0.95-2.5 $\um$, R $\approx$ 120) resolution. All SpeX Prism Library \citep{burgasser2014} L and T dwarfs with R $\approx$ 120 and SNR $>$ 25 were analyzed, mean errors bars at each wavelength were calculated and added to the spectra, and each data point was sampled from a normal distribution characterized by the binned spectral point and error bar at each wavelength to add synthetic noise. The two L2 synthetic data sets and the chemistry profiles used to generate each are shown in Figure \ref{fig:uniformvssc}.

\subsection{Retrieval Framework}\label{retrievalframework}

We adapted the CHIMERA retrieval framework \citep{line2015} for retrieval of hot (T$_{\rm eff}$ = 2000 K) synthetic spectra. We tested four different thermal profile parameterizations and two different chemistry abundance profile parameterizations, with the number of total free parameters ranging from 20 to 36 depending on model parameterization. All models used the radiative transfer core described in \citet{zalesky2022}, which adapted \citet{lacisoinas1991} for use on graphics processing units (GPUs). We utilized the Anaconda Numba \texttt{guvectorize} framework on NVIDIA Quadro RTX 5000 GPUs on Frontera at the Texas Advanced Computing Center. The GPU memory (16 GB) limited the number of simultaneous CPU threads to 4, however Frontera hosts 4 GPUs per computing node, which enabled 4 separate retrievals to be run concurrently. 

All retrievals included the same 14 species (H$_{2}$O, CO, CO$_{2}$, CH$_{4}$, NH$_{3}$, Na, K, PH$_{3}$, TiO, VO, FeH, CrH, MgH, H-) and H$_{2}$-H$_{2}$ and H$_{2}$-He collision induced absorption opacities used to create the synthetic data. Cross sections were sampled at a constant R = 12,500 for SpeX resolution (R = 120), similar to \citet{line2015, zalesky2019,zalesky2022}. 

Parameter estimation for all retrievals was conducted using the \texttt{emcee} package \citep{foremanmackey2013}, as in the previously mentioned studies. All retrievals were run out to 60,000 iterations with 224 to 272 walkers depending on the number of free parameters. The highest and lowest number of parameter models were run out to 120,000 iterations with no significant difference in retrieved posteriors. Initial guesses were constructed by a ``Gaussian-ball'' centered on a by-eye fit for each parameter, and solutions were insensitive to initial guesses. Parameters for all tested retrieval models along with their priors are provided in Table \ref{tab:params}.

\begin{deluxetable*}{llc}
\tablecolumns{3}
\tablecaption{Retrieved Parameters\label{tab:params}}
\tablewidth{0pt}
\tablehead{
\colhead{Parameter} & \colhead{Description}& \colhead{Prior}}
\startdata
log(g) & log of surface gravity [cm s$^{-2}$] & $<$6, M$<$100 M$_{J}$\\
(R/D)$^{2}$ & radius-to-distance scale factor [R$_{J}$/pc] & $<$1, M$<$100 M$_{J}$\\
10$^{b}$ & errorbar inflation & 0.01$\times$min($\sigma_{i}^{2}$), 100$\times$max($\sigma_{i}^{2}$)\\
T$_{i}$\tablenotemark{a} & Temperature [K] at a given pressure level & $<$5500 K \\
$\gamma$, $\beta$\tablenotemark{b} & smoothing hyperparameters (eq. 5, \citet{line2015}) & Inv. Gamma($\Gamma(\gamma,\alpha,\beta$)), $\alpha$=1\\
T$_{\rm app}$\tablenotemark{c} & approximate temperature [K] & $<$4000 K\\
log($\Bar{\kappa}$)\tablenotemark{c} & log of mean opacity & $>$-5,$<$1.5 \\
RCB\tablenotemark{c} & radiative-convective boundary layer [bars] & $>$0.01,$<$30 bars \\
$\gamma_{p}$\tablenotemark{d} & nongreyness & $>$1,$<$100 \\
log($\tau_{lim}$)\tablenotemark{d} & extent of nongreyness & $>$-4,$<$1 \\
log(f$_{i}$) & log of VMR of a uniform gas & $>$-12, $\sum$ f$_{i}$=1\\
BL$_{j}$\tablenotemark{e} & boundary layer [bars] for a nonuniform gas & $>$0.001,$<$315 bars\\
log(a$_{j}$)\tablenotemark{e} & log of VMR of a nonuniform gas below BL$_{j}$ & $>$-12, $\sum$ (f$_{i}$+a$_{j}$)=1 \\
log(b$_{j}$)\tablenotemark{e} & log of VMR of a nonuniform gas above BL$_{j}$ & $>$-12, $\sum$ (f$_{i}$+b$_{j}$)=1\\
\enddata
\tablecomments{$^{a}$smoothed and unsmoothed PT retrievals only; $^{b}$smoothed PT retrievals only; $^{c}$grey and nongrey RC PT retrievals only; $^{d}$nongrey RC PT retrievals only; $^{e}$nonuniform gas retrievals only.}
\end{deluxetable*}
\subsubsection{Tested Thermal Profiles}
The thermal profile refers to how the temperature varies with pressure in the atmosphere. In this work, 4 different thermal profile parameterizations were tested to determine the accuracy with which each captures the ground truth PT profile and ground truth bulk properties. Each of these parameterizations are tested on the uniform chemistry spectrum.
\begin{enumerate}
    \item ``Unsmoothed'' Free. The most flexible thermal profile is the free-unsmoothed profile (henceforth ``unsmoothed'' PT profile), which is parameterized by 15 temperature variables equidistant in log(P) space between -3.5 and 2.5. This 15 point profile is connected by Hermite spline interpolation and is then interpolated onto the finer 70 point pressure grid for the radiative-transfer calculation. There is no smoothing, which can allow for unphysical oscillations from layer to layer.
    \item ``Smoothed'' Free. The next most flexible profile is the free-smoothed profile (``smoothed'' PT profile) from \citet{line2015}, which is parameterized as above, but with the addition of 2 smoothing hyper-parameters that penalize the second derivative of the profile. This profile allows oscillations if there is sufficient evidence to justify curvature of the profile. This profile becomes less adequate as the probed pressure layers narrow for hotter objects.
    \item  ``Grey RC''. We constructed a third radiative-convective profile characterized by a gray-radiative upper atmosphere and a convective adiabat lower atmosphere (``grey RC'' profile). This profile is parameterized by 4 parameters. The gray-radiative portion from \citet{guillot2010} is constructed by surface gravity (log(g)) and mean opacity (log($\Bar{\kappa}$)) parameters to control the optical depth to pressure mapping and an approximate temperature parameter (T$_{\rm app}$) to set the Eddington approximation grey-radiative profile, as described in Equation \ref{eq:eddington}.
    \begin{equation}
    T^{4}(\tau) = \frac{3 T^{4}_{app}}{4} \left( \frac{2}{3} + \tau \right)
    \label{eq:eddington}
    \end{equation}
    The final parameter is the location of the radiative-convective boundary (RCB) in bars. The convective part of the atmosphere is set by a dry adiabat as described in \citet{robinsoncatling2012} and characterized by the temperature and pressure at the boundary and is set by $\frac{\partial \ln{T}}{\partial \ln{P}}$ of the final two radiative layers. The RCB can vary within the prior because the PT profile is parameterized independently from the chemistry, unlike \citet{molliere2020} where the RCB and convective adiabat are determined by the equilibrium chemistry grid.
    \item ``Nongrey RC''. The grey-radiative profile from \citet{guillot2010} causes an isothermal upper atmosphere that is hotter than the thermal profiles from the self-consistent Sonora models. We constructed a fourth PT profile to incorporate the non-grey effects in the upper atmosphere (``nongrey RC'' profile). We use the non-irradiated, non-grey parameterization from \citet{chandrasekhar1935} as described by equation 28 in \citet{parmentierguillot2014} below.
    \begin{eqnarray}\nonumber
    T^{4}(\tau) = \frac{3 T^{4}_{app}}{4} \left( \tau + \frac{\frac{2}{3} + \sqrt{\frac{1}{3\gamma_{P}}}}{1 + \frac{1}{2}\sqrt{3\gamma_{P}}} \right)  +
    \end{eqnarray}
    \begin{eqnarray}
    \frac{3 T^{4}_{app}}{4} \left( \frac{\gamma_{P} - 1}{\sqrt{\gamma_{P}}} \right)\frac{\frac{1}{\sqrt{3}}+\sqrt{\gamma_{P}}\tau_{lim}}{1+\frac{1}{2}\sqrt{3\gamma_{P}}} \left( 1 - e^{\frac{-\tau}{\tau_{lim}}} \right)
    \label{eq:rc}
    \end{eqnarray}
    This profile is constructed with 6 parameters: log(g), log($\Bar{\kappa}$), T$_{\rm int}$, the radiative-convective boundary (RCB), $\tau_{lim}$, $\gamma_{P}$. The lower atmosphere is characterized by a convective adiabat as described above, but the upper atmosphere is constrained by Equation \ref{eq:rc}, where $\gamma_{P}$ controls the non-greyness of the upper atmosphere and $\tau_{lim}$ controls the optical depth to which the nongreyness extends in the atmosphere. Like the grey radiative-convective profile, this profile is also parameterized independently from the chemistry.
\end{enumerate}
 
\subsubsection{Tested Chemical Abundance Profiles}
The chemical abundance profiles of all included opacity species throughout the atmosphere are described by volume mixing ratios (VMRs). We tested two free chemistry models. 

The first chemical abundance profile parameterization tested was the existing free chemistry model which used uniform-with-pressure profiles for all 14 gases (H$_{2}$O, CO, CO$_{2}$, CH$_{4}$, NH$_{3}$, Na, K, PH$_{3}$, TiO, VO, FeH, CrH, MgH, H-) retrieved. The chemistry in these models was controlled by 14 mixing ratios, one for each gas. This provided the flexibility to retrieve disequilibrium species, but did not allow species to vary throughout the atmosphere. This parameterization was used in all retrievals on the uniform chemistry spectrum and also tested on the self-consistent chemistry spectrum.

The second type of chemical profile tested was a step function, where the abundance profile for a single gas was controlled by 3 parameters, an upper mixing ratio, a lower mixing ratio, and a pressure layer where the abundance changed (`nonuniform' profile). For these retrievals, we set all profiles as uniform with the exception of 1 gas (and later 2 gases), which had the nonuniform profile. The chemistry in these models was controlled by 16 parameters (13 uniform profiles and 1 nonuniform profile). This chemistry parameterization was tested on the self-consistent chemistry spectrum.

A detailed description of each retrieval test performed including the spectrum retrieved on, the thermal profile parameterization, the chemistry parameterization, and the total number of parameters is provided in Table \ref{tab:retrievaltests}.

\begin{deluxetable}{cccc}
\tablecolumns{4}
\tablecaption{Retrieval Tests\label{tab:retrievaltests}}
\tablewidth{0pt}
\tablehead{
\colhead{Data} & \colhead{PT Profile}& \colhead{Nonuniform Gas}& \colhead{Parameters}}
\startdata
Uniform & Grey RC &  None & 20  \\
Uniform & Nongrey RC & None & 22 \\
Uniform & Unsmoothed & None & 32  \\
Uniform & Smoothed & None & 34  \\
Self-Consistent & Grey RC &  None & 20  \\
Self-Consistent & Nongrey RC & None & 22 \\
Self-Consistent & Unsmoothed & None & 32  \\
Self-Consistent & Smoothed & None & 34 \\
Self-Consistent & Unsmoothed & FeH & 34  \\
Self-Consistent & Unsmoothed & K & 34  \\
Self-Consistent & Unsmoothed & H$_{2}$O & 34  \\
Self-Consistent & Unsmoothed & VO & 34  \\
Self-Consistent & Unsmoothed & CrH & 34  \\
Self-Consistent & Unsmoothed & CO & 34 \\
Self-Consistent & Unsmoothed & TiO & 34  \\
Self-Consistent & Unsmoothed & Na & 34 \\
Self-Consistent & Unsmoothed & FeH + K & 36  \\
\enddata
\end{deluxetable}
\begin{deluxetable*}{ccccccccc}
\tablecolumns{9}
\tablecaption{Thermal Profile Retrieval Results\label{tab:PTresults}}
\tablewidth{0pt}
\tablehead{
\colhead{Data} & \colhead{PT Profile} & \colhead{log(g)}& \colhead{Radius} & \colhead{Mass} & \colhead{T$_{\rm eff}$} & \colhead{[M/H]} & \colhead{C/O} & \colhead{$\Delta$BIC}}
\startdata
Uniform & Sonora &  5.0 & 1.084 & 47.4 & 2124 & -0.015 & 0.487 & N/A \\
Uniform & Grey RC &  5.04$\pm$0.06 & 1.09$\pm$0.01 & 52$^{+7}_{-6}$ & 2119$^{+8}_{-7}$ & \textbf{0.03}$^{+0.05}_{-0.03}$ & 0.50$\pm$0.01 & -16 \\
Uniform & Nongrey RC & \textbf{5.14}$^{+0.06}_{-0.04}$ & \textbf{1.07}$\pm$0.01 & \textbf{65}$^{+9}_{-5}$ & \textbf{2132}$\pm$6 & \textbf{0.07}$^{+0.04}_{-0.02}$ & 0.49$\pm$0.01 & -8 \\
Uniform & Unsmoothed & 4.99$^{+0.03}_{-0.04}$ & 1.08$\pm$0.01 & 46$^{+4}_{-3}$  & 2124$\pm$5 & -0.02$\pm$0.02 & 0.48$\pm$0.01& - \\
Uniform & Smoothed & 4.98$\pm$0.03 & 1.09$\pm$0.01 & 46$\pm$3 & 2121$\pm$4 & -0.02$\pm$0.02 & 0.49$\pm$0.01 & -172 \\
\hline
Self-Consistent & Sonora &  5.0 & 1.084 & 47.4 & 2038 & 0.00 & 0.534 & N/A \\
Self-Consistent & Grey RC &  \textbf{5.27}$\pm$0.03 & 1.13$^{+0.02}_{-0.07}$ & \textbf{96}$^{+3}_{-7}$ & 1987$^{+70}_{-25}$ & \textbf{0.49}$^{+0.50}_{-0.25}$ & \textbf{0.65}$^{+0.04}_{-0.06}$ & -23 \\
Self-Consistent & Nongrey RC & 5.25$^{+0.05}_{-0.43}$ & 1.11$\pm$0.03 & 88$^{+8}_{-57}$ & \textbf{2007}$^{+29}_{-23}$ & \textbf{0.19}$^{+0.23}_{-0.11}$ & 0.59$^{+0.04}_{-0.06}$ & -13 \\
Self-Consistent & Unsmoothed & \textbf{5.20}$^{+0.08}_{-0.09}$ & 1.08$\pm$0.02 & \textbf{75}$^{+14}_{-13}$ & 2038$^{+18}_{-20}$ & \textbf{0.16}$^{+0.12}_{-0.07}$ & 0.56$^{+0.04}_{-0.05}$ & - \\
Self-Consistent & Smoothed & \textbf{5.25}$^{+0.04}_{-0.11}$ & \textbf{1.11}$\pm$0.02 & \textbf{90}$^{+8}_{-17}$ & \textbf{2007}$^{+23}_{-24}$ & \textbf{0.17}$^{+0.12}_{-0.10}$ & 0.59$^{+0.04}_{-0.06}$ & -219 \\
\enddata
\end{deluxetable*}
\begin{deluxetable*}{ccccccccc}
\tablecolumns{9}
\tablecaption{Self-Consistent Chemistry Retrieval Results\label{tab:gasresults}}
\tablewidth{0pt}
\tablehead{
\colhead{Data} & \colhead{Nonuniform Gas} & \colhead{log(g)}& \colhead{Radius} & \colhead{Mass} & \colhead{T$_{\rm eff}$} & \colhead{[M/H]} & \colhead{C/O} & \colhead{$\Delta$BIC}}
\startdata
Self-Consistent & Sonora &  5.0 & 1.084 & 47.4 & 2038 & 0.00 & 0.534 & N/A \\
Self-Consistent & None &  \textbf{5.20}$^{+0.08}_{-0.09}$ & 1.08$\pm$0.02 & \textbf{75}$^{+14}_{-13}$ & 2038$^{+18}_{-20}$ & \textbf{0.16}$^{+0.12}_{-0.07}$ & 0.56$^{+0.04}_{-0.05}$ & - \\
Self-Consistent & FeH & 5.06$^{+0.05}_{-0.06}$ & 1.08$\pm$0.01 & 54$^{+6}_{-7}$ & 2040$\pm$7 & 0.013$\pm$0.03 & 0.53$\pm$0.01 & 289 \\
Self-Consistent & K & \textbf{5.24}$^{+0.06}_{-0.09}$ & 1.08$^{+0.02}_{-0.01}$ & \textbf{81}$^{+10}_{-13}$ & 2039$^{+14}_{-15}$ & \textbf{-0.30}$^{+0.19}_{-0.09}$ & \textbf{0.15}$^{+0.04}_{-0.09}$ & 35 \\
Self-Consistent & H$_{2}$O & 5.03$^{+0.15}_{-0.35}$ & 1.10$^{+0.03}_{-0.02}$ & 53$^{+19}_{-28}$ & 2021$^{+23}_{-29}$ & \textbf{0.22}$^{+0.25}_{-0.18}$ & \textbf{0.33}$^{+0.11}_{-0.13}$ & 35 \\
Self-Consistent & VO & \textbf{5.19}$^{+0.09}_{-0.12}$ & 1.08$\pm$0.02 & \textbf{73}$\pm$17 & 2034$^{+17}_{-19}$ & 0.09$^{+0.11}_{-0.10}$ & 0.56$\pm$0.04 & 5.7 \\
Self-Consistent & CrH & \textbf{5.22}$^{+0.08}_{-0.09}$ & 1.08$^{+0.2}_{-0.1}$  & \textbf{77}$^{+13}_{-14}$ & 2039$^{+16}_{-17}$ & \textbf{0.11}$^{+0.11}_{-0.08}$ & 0.56$\pm$0.04 & 3.7 \\
Self-Consistent & CO & \textbf{5.21}$\pm$0.08 & 1.08$\pm$0.02 & \textbf{76}$^{+14}_{-11}$ & 2039$^{+15}_{-18}$ & -0.07$^{+0.28}_{-0.30}$ & 0.34$^{+0.25}_{-0.30}$ & 0.56 \\
Self-Consistent & TiO & \textbf{5.21}$^{+0.08}_{-0.11}$ & 1.08$\pm$0.02 & \textbf{77}$\pm$14 & 2039$^{+13}_{-22}$ & \textbf{0.10}$^{+0.11}_{-0.08}$ & 0.56$\pm$0.04 & -0.48 \\
Self-Consistent & Na & \textbf{5.19}$^{+0.09}_{-0.11}$ & 1.08$^{+0.02}_{-0.01}$ & \textbf{74}$\pm$15 & 2037$^{+15}_{-20}$ & 0.09$^{+0.09}_{-0.10}$ & 0.56$\pm$0.04 & -3.1 \\
Self-Consistent & FeH + K & 5.02$\pm$0.07 & 1.08$\pm$0.09 & 50$^{+8}_{-7}$ & 2034$^{+8}_{-9}$ & \textbf{-0.06}$\pm$0.03 & 0.54$\pm$0.02 & 279 \\
\enddata
\end{deluxetable*}

\section{Results}\label{Results}
In this section we present results from the retrieval tests. Comparisons of retrieved thermal profiles, abundance profiles, and bulk properties are presented. While some properties like surface gravity, radius, and abundance profiles are retrieved directly, other bulk properties like PT profiles, effective temperatures, metallicity, and C/O are derived based on retrieved parameters. The derivations of each of these parameters are described below.

Free (smoothed and unsmoothed) PT profiles are constructed by using Hermite spline interpolation between the 15 directly retrieved temperature points. Radiative-convective profiles are constructed using the retrieved parameters and equations \ref{eq:eddington} or \ref{eq:rc}. As in \citet{line2017}, \citet{zalesky2019}, and \citet{zalesky2022}, effective temperature is derived by equating Boltzmann's law to the bolometric fluxes between 0.7 and 20 $\um$ derived from 1000 model spectra drawn from the posterior.

Metallicity is computed as, 
\begin{equation}
\left[ M/H \right] = log\left( \frac{(M/H)_{retrieved}}{(M/H)_{solar}}\right)
\label{eq:metallicity}
\end{equation}
where the retrieved metallicity is taken to be the summation of the elemental species included in the retrieval model. The C/O ratio is computed as, 
\begin{equation}
\frac{C}{O} = \frac{\sum C}{\sum O} \approx \frac{CH_{4} + CO + CO_{2}}{H_{2}O + CO + 2CO_{2} + VO + TiO}
\label{eq:coratio}
\end{equation}
\begin{figure*}[!htp]
    \centering
    \includegraphics[width=0.52\linewidth]{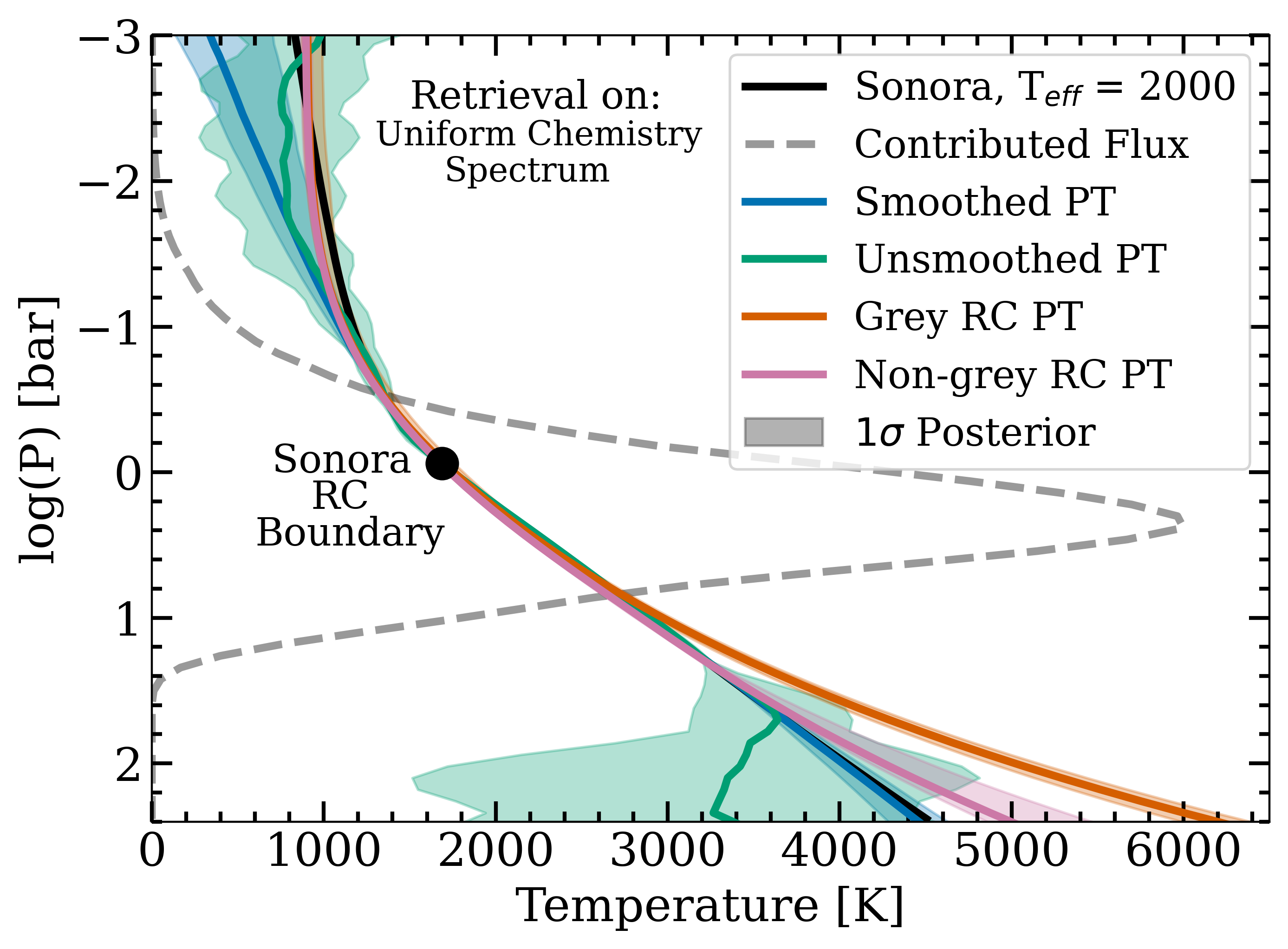}
    \includegraphics[width=0.455\linewidth]{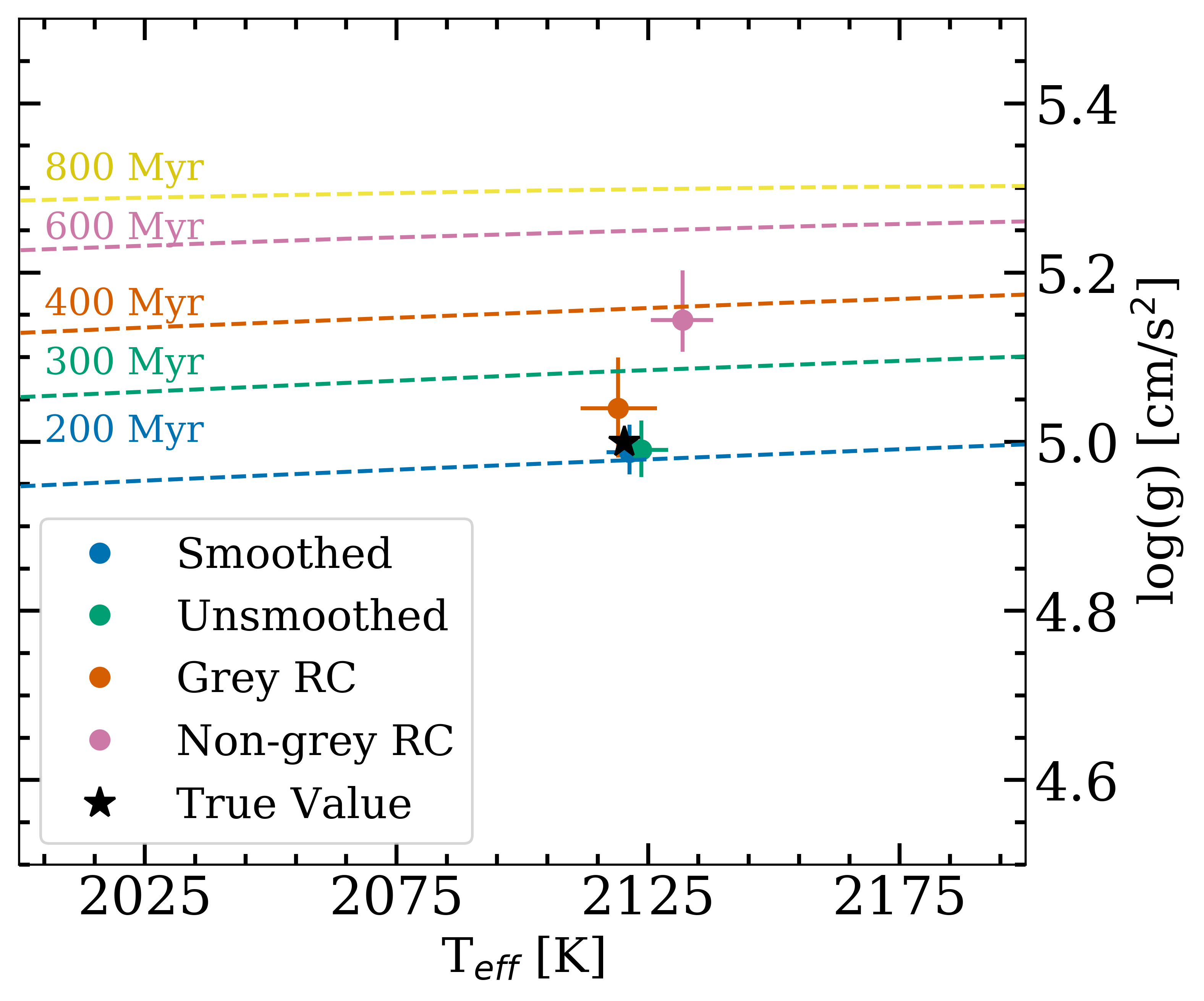}
    \includegraphics[width=1.0\linewidth]{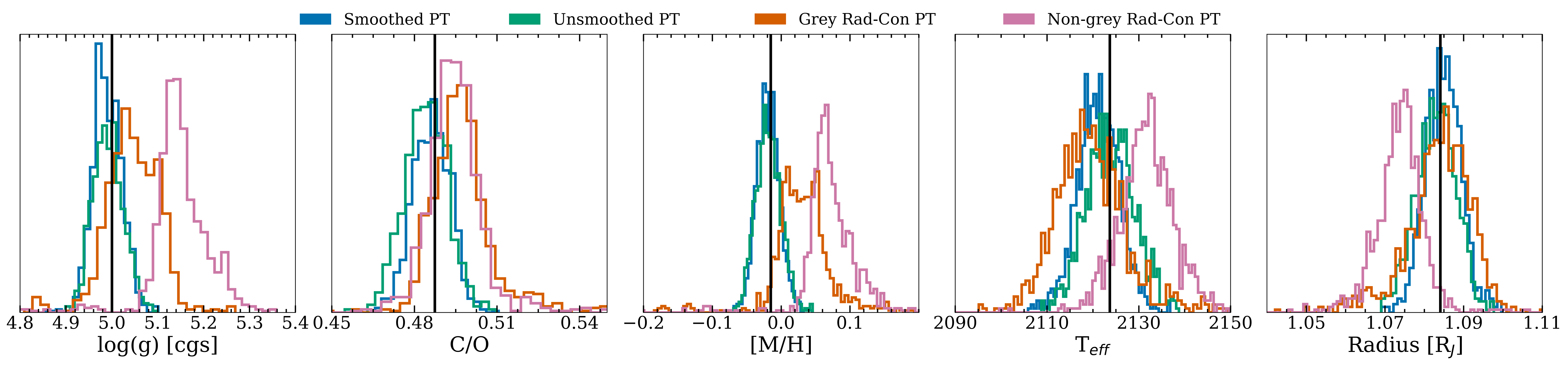}
    \caption{Retrieved PT profiles, bulk properties, and inferred ages using 4 PT profile parameterizations (colors), ground truth Sonora values (black), and isochrones from \citet{marley2021ApJ...920...85M} (dashed colors). Retrievals done on the uniform chemistry spectrum. The free thermal profiles accurately retrieve bulk properties but do not reliably constrain the upper atmosphere.}
    \label{fig:toyptparams}
\end{figure*}
\begin{figure*}[!htp]
    \centering
    \includegraphics[width=1.0\linewidth]{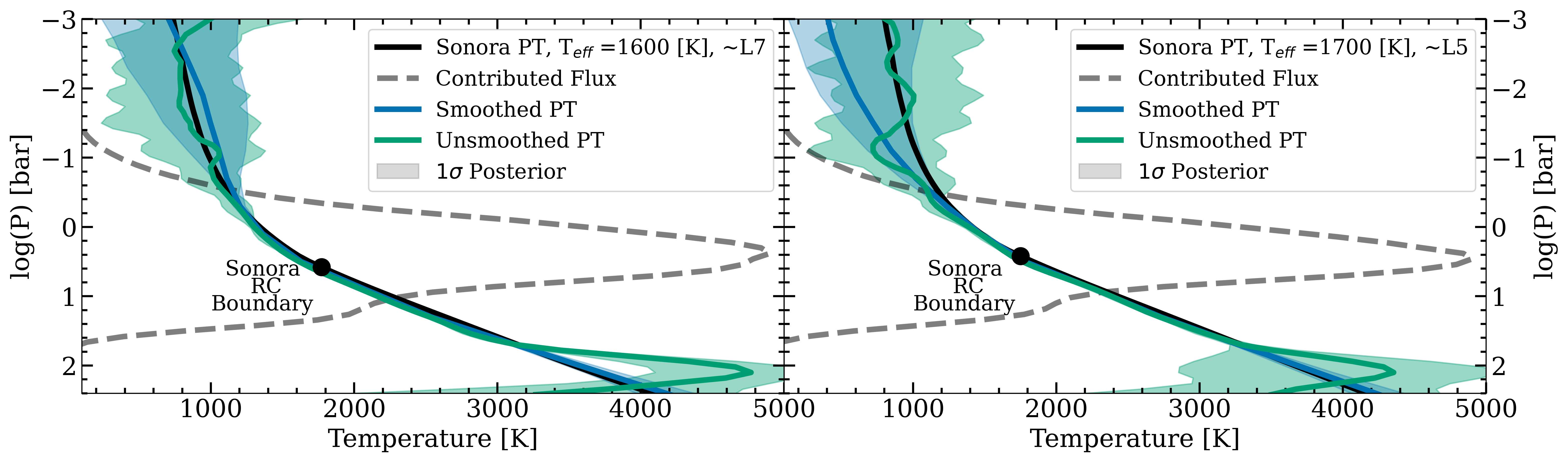}
    \caption{Retrieved thermal profiles for two synthetic L dwarfs showing that a smoothed PT profile can accurately retrieve the thermal profile of objects later than L6, but retrieves a cooler upper atmosphere for objects earlier than L5.}
    \label{fig:PT1600_1700}
\end{figure*}
More complex thermal and chemistry profile parameterizations use more parameters but may provide a better fit to the data. The Bayesian Information Criterion (BIC) was computed for all retrievals to determine if the inclusion of additional parameters resulted in better fits to the data. BIC was calculated as,

\begin{equation}
BIC = -2 \ln{(L)}+ \ln{(N)} K
\label{eq:deltabic}
\end{equation}

where ln(L) is the log-likelihood of the best fit model, N is the number of data points and K is the number parameters. The model with the lowest BIC signals the better model. As in \citet{kassraftery1995} and \citet{gonzales2021L3}, we use the following intervals for selecting between two models with evidence against the higher BIC as: 0 $<$ $\Delta$BIC $<$ 2: no preference worth mentioning; 2 $<$ $\Delta$BIC $<$ 6: positive; 6 $<$ $\Delta$BIC $<$ 10: strong; and 10 $<$ $\Delta$BIC: very strong.

\subsection{Thermal Profile Retrievals} \label{sec:thermalprofilesresults}
All PT profile parameterizations were tested on the uniform chemistry spectrum. Figure \ref{fig:toyptparams} shows the median and 1$\sigma$ PT profiles retrieved by each model compared to the ground truth Sonora PT profile for the L2 object. The bottom figure shows bulk properties retrieved or derived by each model compared to ground truth Sonora bulk properties. Also shown are the inferred ages based on retrieved surface gravity and effective temperature. Table \ref{tab:PTresults} shows the retrieved bulk properties retrieved or derived for the four PT profiles tested. Bolded values indicate disagreement with the ground truth values. Figure \ref{fig:uniabund_append} in Appendix \ref{AppendixAA} shows all retrieved chemical abundances for the four PT parameterizations. Only the smoothed and unsmoothed PT profiles accurately retrieve all abundances and bulk properties from the uniform chemistry spectrum. 

NIR SpeX resolution spectra for objects of this type probe a narrow pressure range, as shown by the grey dashed line in Figure \ref{fig:toyptparams}. The pressures probed contain mostly a linear part of the temperature profile, so a more linear PT profile is preferred in a smoothed parameterization. The unsmoothed PT profile is only constrained in this probed region. Table \ref{tab:PTresults} shows that neither the smoothed nor unsmoothed PT profiles bias retrieved bulk properties, but Figure \ref{fig:toyptparams} shows that a smoothed PT profile retrieves a colder profile above the photosphere while an unsmoothed PT profile retrieves unphysical oscillations above and below the photosphere for an $\sim$L2 object with  T$_{\rm eff}$ = 2000 K. 

\subsubsection{Thermal Profile Retrievals on Different Spectral Types}
We conducted additional retrievals on uniform chemistry data for an $\sim$L5 object with T$_{\rm eff}$ = 1700 K and an $\sim$L7 object with T$_{\rm eff}$ = 1600 K to determine for which spectral type a retrieved smoothed PT profile can no longer be trusted above the photosphere. We retrieved both objects with a smoothed and unsmoothed PT profile parameterization and the retrieved PT profiles can be seen in Figure \ref{fig:PT1600_1700}. The smoothed PT profile retrieves an accurate profile for L6 objects and later but a cooler upper atmosphere for objects earlier than L5. 
\begin{figure*}[!htp]
    \centering
    \includegraphics[width=0.52\linewidth]{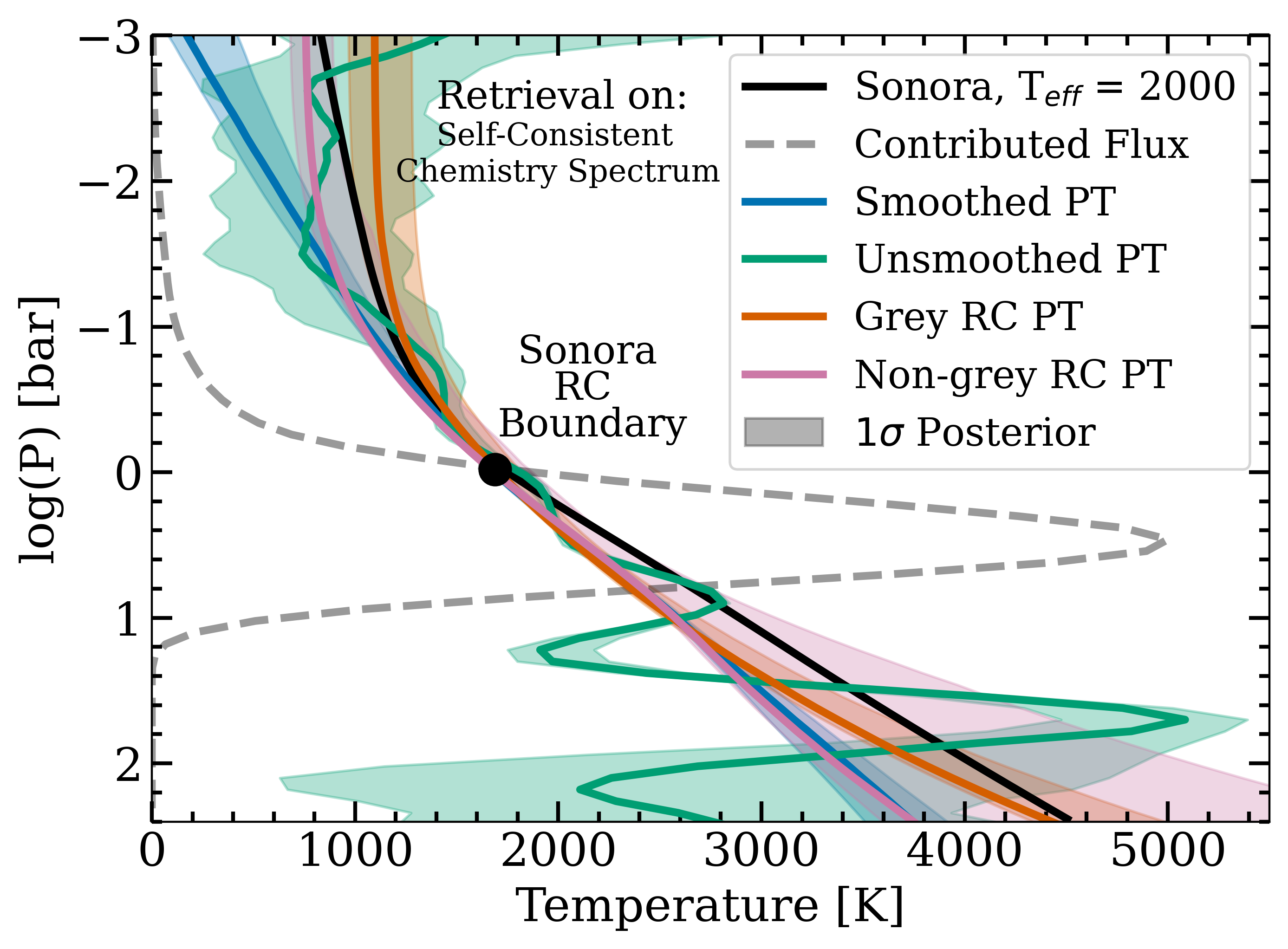}
    \includegraphics[width=0.455\linewidth]{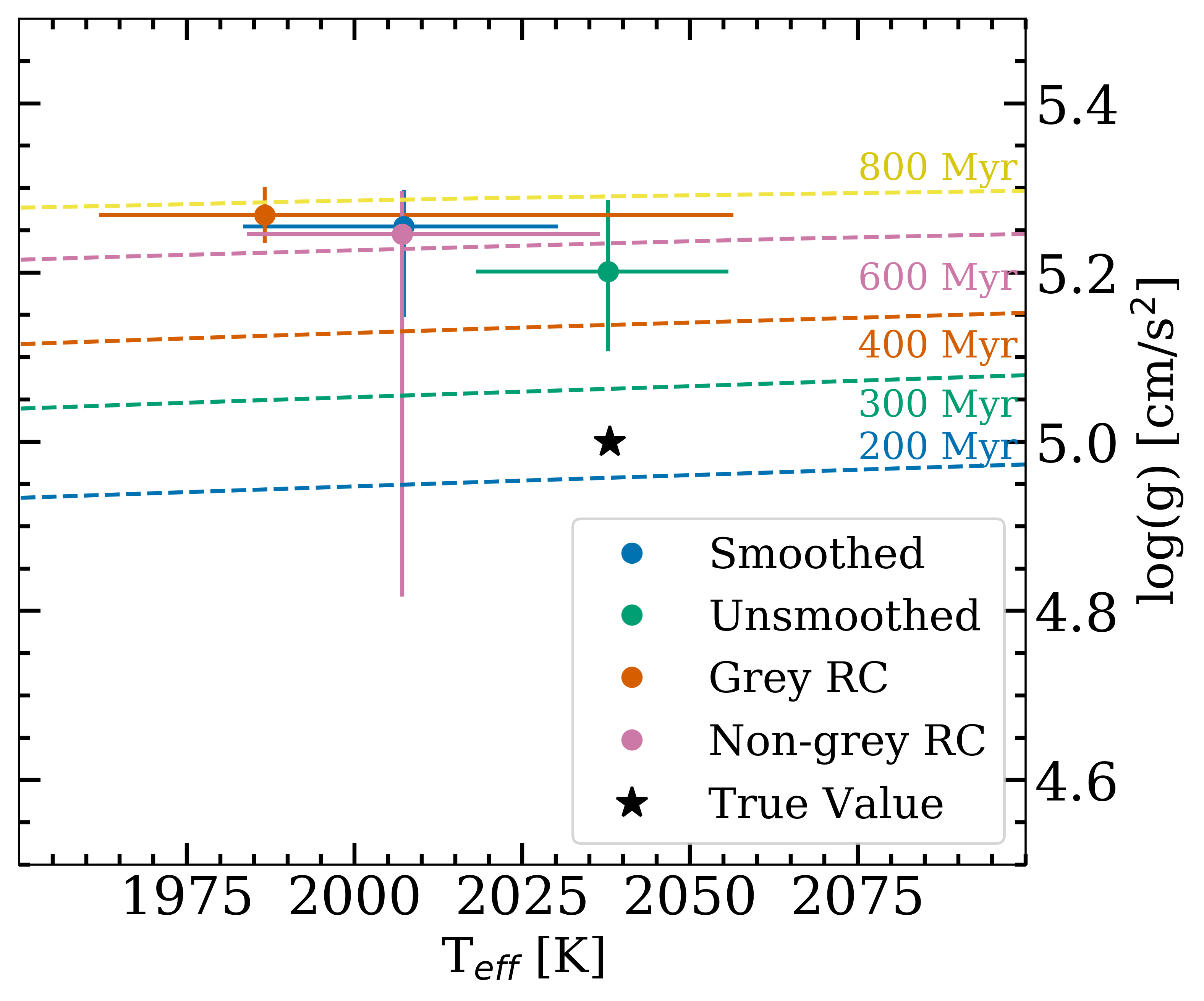}
    \includegraphics[width=1.0\linewidth]{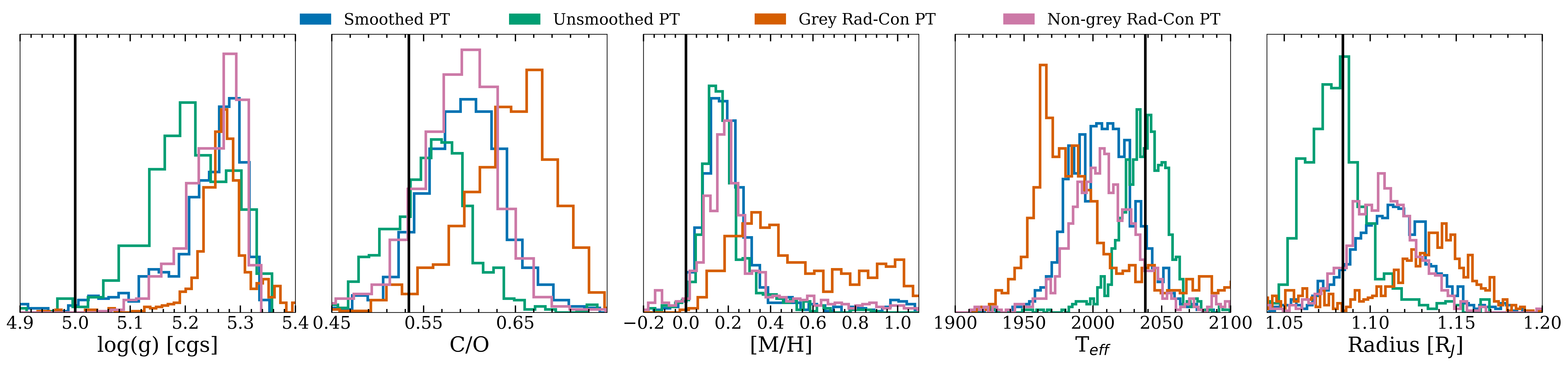}
    \caption{Same as Figure \ref{fig:toyptparams} but for retrievals done with the self-consistent chemistry spectrum. Thermal profiles, bulk properties, and inferred ages, are inaccurately retrieved regardless of PT profile parameterization.}
   \label{fig:nontoyptparams}
\end{figure*}
\begin{figure*}[!tb]
    \centering
    \includegraphics[width=0.52\linewidth]{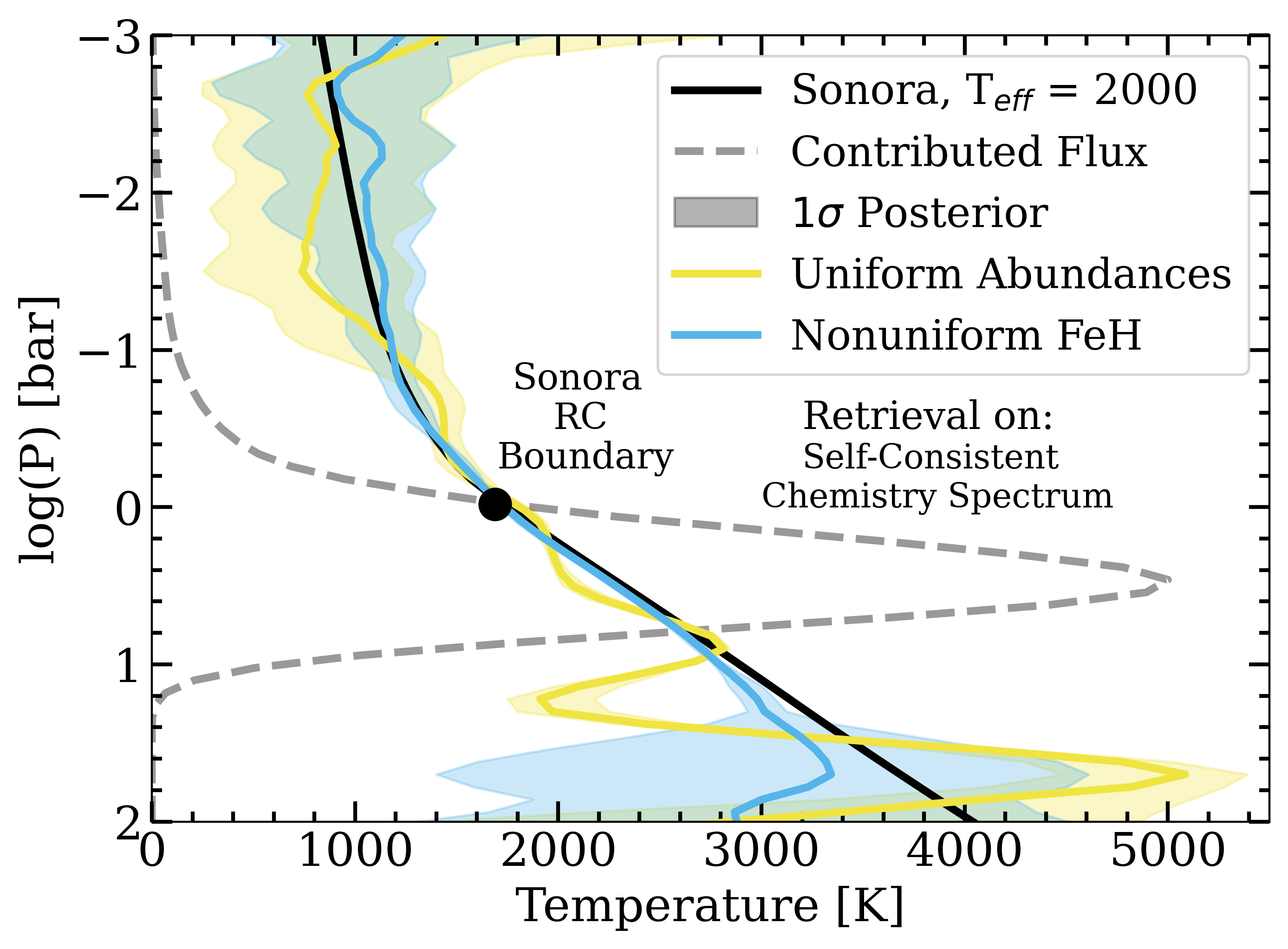}
    \includegraphics[width=0.455\linewidth]{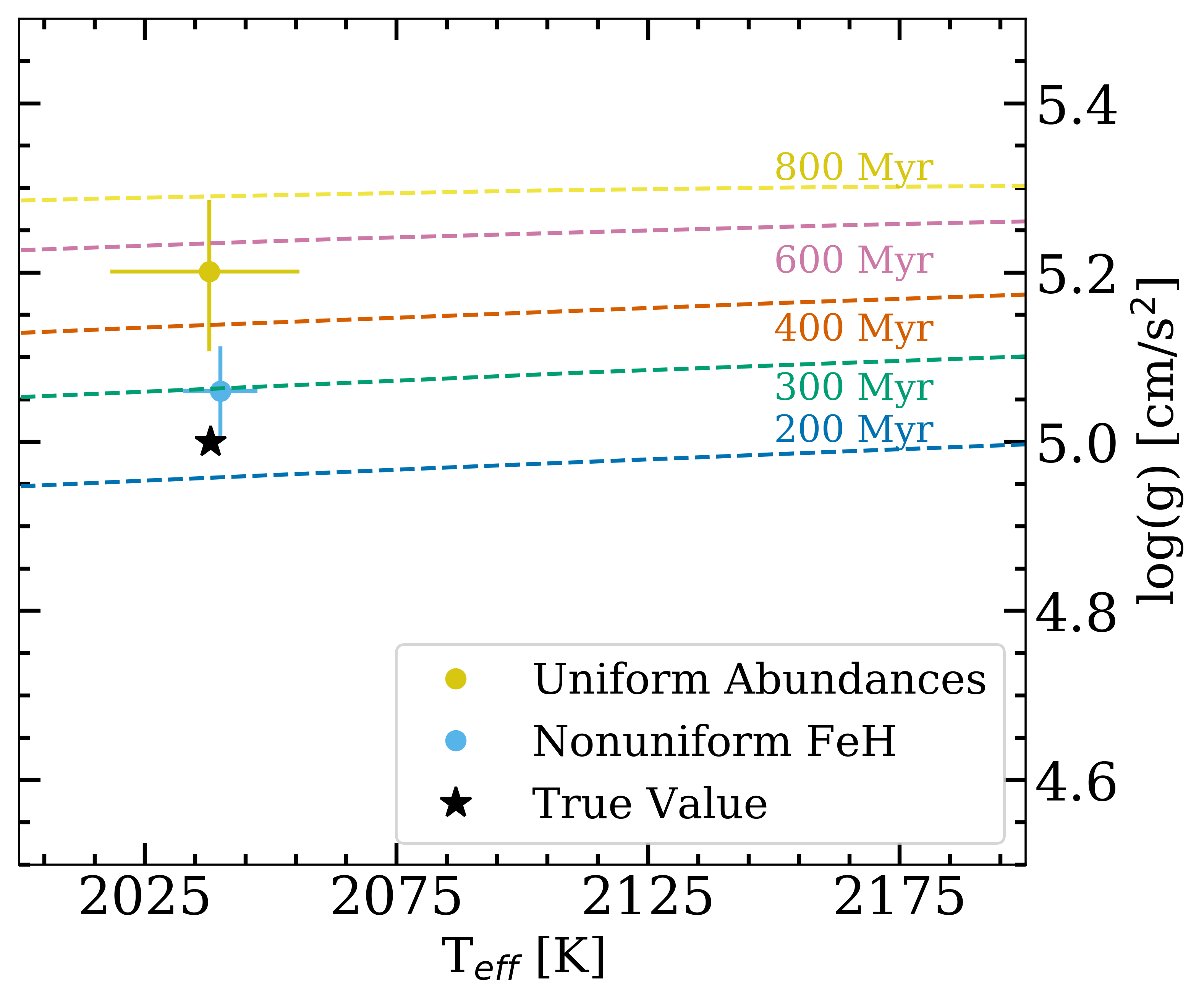}
    \includegraphics[width=1.0\linewidth]{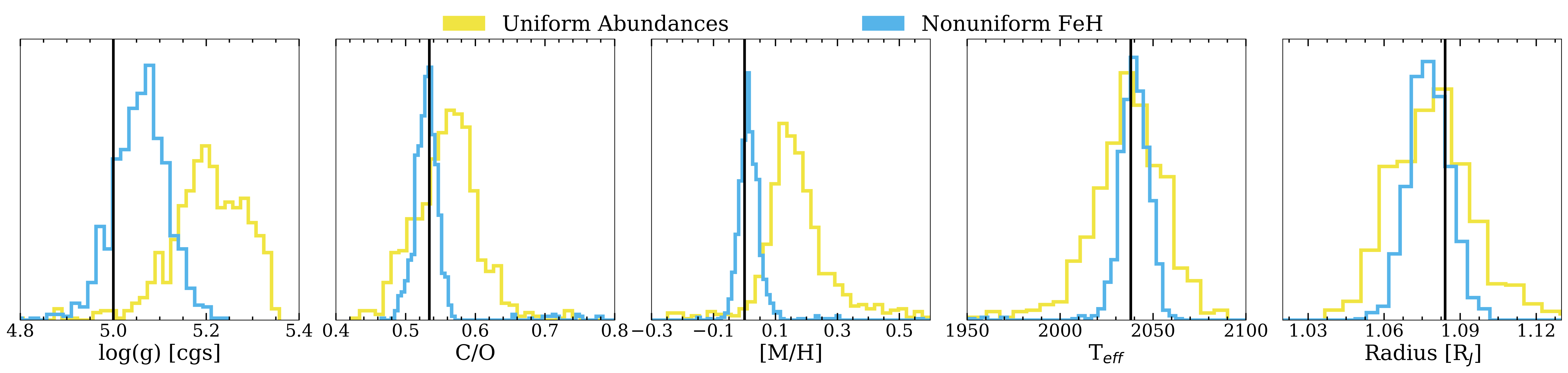}
    \caption{Retrieved PT profiles, bulk properties, and inferred ages using 2 different free chemistry parameterizations (colors), ground truth Sonora values (black), and isochrones from \citet{marley2021ApJ...920...85M} (dashed colors). Accurate bulk properties are only retrieved by the including a nonuniform abundance profile for iron hydride (FeH).}
    \label{fig:fehptparams}
\end{figure*}

\subsection{Thermal Profile Retrievals on the Self-Consistent Chemistry Spectrum}
Most brown dwarf retrievals assume uniform-with-altitude abundance \citep{line2017, zalesky2019, burningham2017}; we mimicked this approach for our synthetic data by retrieving on the self-consistent chemistry spectrum with all four PT profile parameterizations assuming uniform chemistry parameterizations. The retrieved PT profiles, bulk properties, and inferred ages are shown in Figure \ref{fig:nontoyptparams}, and the bulk properties retrieved and derived are shown in Table \ref{tab:PTresults}. The smoothed, unsmoothed, grey, and nongrey radiative convective profiles do not agree with the Sonora PT profile, and almost all retrieved or derived bulk properties and ages for all PT parameterizations are in disagreement at the 1$\sigma$ level. Figure \ref{fig:nonuniabund_append} in Appendix \ref{AppendixAA} shows the retrieved abundances for all four PT parameterizations. All four models retrieve a higher CO abundance than the input profile, and all models except the unsmoothed model retrieve higher CO$_{2}$ and H- abundances. For this spectral type, no choice of PT profile can accurately retrieve values in this regime if uniform chemistry is used. 

\subsection{Chemical Abundance Profile Retrievals}
All retrievals testing different abundance profile parameterizations were conducted on the self-consistent chemistry spectrum, which used both the Sonora PT profile and abundance profiles. All retrievals used the unsmoothed PT profile parameterization, which provided the most unbiased results in Section \ref{sec:thermalprofilesresults}.

We include one fully uniform abundance retrieval, in which all 14 gases had uniform (constant with altitude) abundances, and eight retrievals in which 13 gases had uniform abundances and 1 gas had the nonuniform abundance parameterization (step function). The six gases that were not tested for nonuniform profiles were the gases for which only upper limits were retrieved in the retrievals on the uniform chemistry spectrum, indicating that not enough information in the spectrum exist to justify a more complex treatment. An additional retrieval with both nonuniform FeH and K was also performed.

Inclusion of a nonuniform gas in the retrieval adds an additional 2 parameters to the model (secondary abundance and boundary pressure). Retrieved and derived bulk properties for all 10 retrievals are shown in Table \ref{tab:gasresults}. We only retrieve accurate bulk properties when nonuniform FeH is included. The nonuniform FeH retrieval was strongly preferred over a fully uniform retrieval ($\Delta$BIC = 289) and was also strongly preferred over every other retrieval. 

The two most preferred nonuniform gas retrievals were FeH ($\Delta$BIC = 289) and K ($\Delta$BIC = 35). An additional retrieval with both nonuniform FeH and nonuniform K was strongly preferred over the uniform chemistry model ($\Delta$BIC = 279), but was not preferred over nonuniform FeH alone ($\Delta$BIC = -11), indicating that adding additional nonuniform gases beyond FeH was not justified.

The retrieved PT profiles of the fiducial fully uniform chemistry retrieval and the winning nonuniform FeH retrieval are shown in Figure \ref{fig:fehptparams}. The only retrieval in which the PT profile is retrieved accurately through the photosphere is the nonuniform FeH retrieval. While a uniform chemistry retrieval was able to accurately retrieve the PT profile in the photosphere of the uniform chemistry spectrum (Figure \ref{fig:toyptparams}), it fails on the self-consistent chemistry spectrum. The retrieved profile shows many nonphysical oscillations. In particular, the cooler-than-expected oscillation at log(P) = 1.2 and the warmer-than-expected oscillation at log(P) = 1.8 also appear in all retrievals that have uniform FeH parameterizations, regardless of the other nonuniform gases.

Figure \ref{fig:fehptparams} also shows the retrieved or derived bulk properties including surface gravity, radius, C/O ratio, metallicity, and effective temperature, along with inferred ages and isochrones from \citet{marley2021ApJ...920...85M}. The fully uniform free chemistry retrieval biased retrieved surface gravity, mass, and metallicity. The age inferred from surface gravity and effective temperature was not in agreement with the ground truth value at the 1$\sigma$ level.

\begin{figure}[!htb]
    \centering
    \includegraphics[width=0.48\textwidth]{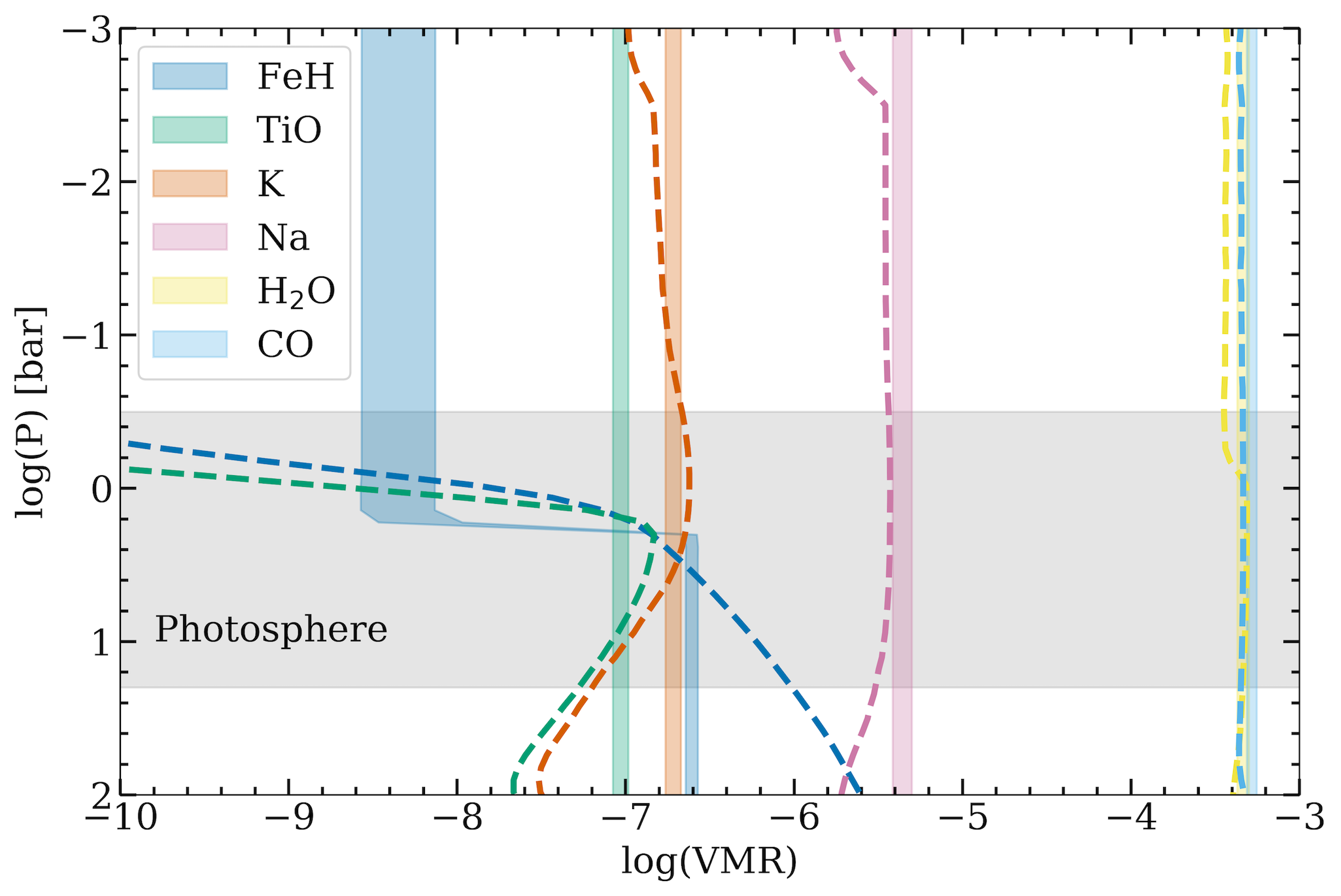}
    \caption{Sonora (dashed) and 1$\sigma$ retrieved (shaded) abundance profiles of the 6 most abundant species with the photosphere shaded in grey. All gas species except for FeH are retrieved as uniform-with-pressure. FeH is retrieved as a step function that accurately retrieves rainout information.}
    \label{fig:abundfeh}
\end{figure}

The retrieved chemistry abundance profiles from the winning nonuniform FeH retrieval is shown in Figure \ref{fig:abundfeh}. Only the 6 most abundant gases are shown, and all gases except FeH have uniform retrieved abundances. The Sonora abundance profiles for prominent species like CO, H$_{2}$O, Na, and K have relatively uniform abundances that are closely matched by the uniform parameterization. TiO has a strongly nonuniform profile, and the retrieved uniform abundance matches the expected abundance through the photosphere (shaded grey). The FeH abundance profile (deep blue) is a step function that tracks the expected abundances of more FeH in the lower atmosphere followed by swift removal from the atmosphere around 1 bar due to rainout. All retrieved abundances for this retrieval are shown in Figure \ref{fig:nonuniabund_append} in Appendix \ref{AppendixAA}. The corner plot and retrieved spectra are provided in Appendix \ref{AppendixA}.
\section{Discussion}\label{Discussion}
We performed retrievals on synthetic L dwarf spectra for which we know all ground truth values in order to answer two questions. First, which PT profile parameterizations accurately retrieve bulk properties and thermal profiles with spectral information from limited pressure ranges? Second, what kind of free chemistry profile parameterizations accurately retrieve bulk properties in objects with abundances that are strongly nonuniform? In this section we discuss the answers to these questions and provide additional analyses that aim to inform future L dwarf retrievals.

\subsection{Constraining Thermal Profiles in L Dwarf Atmospheres}
\begin{figure*}[!htb]
    \centering
    \includegraphics[width=0.47\linewidth]{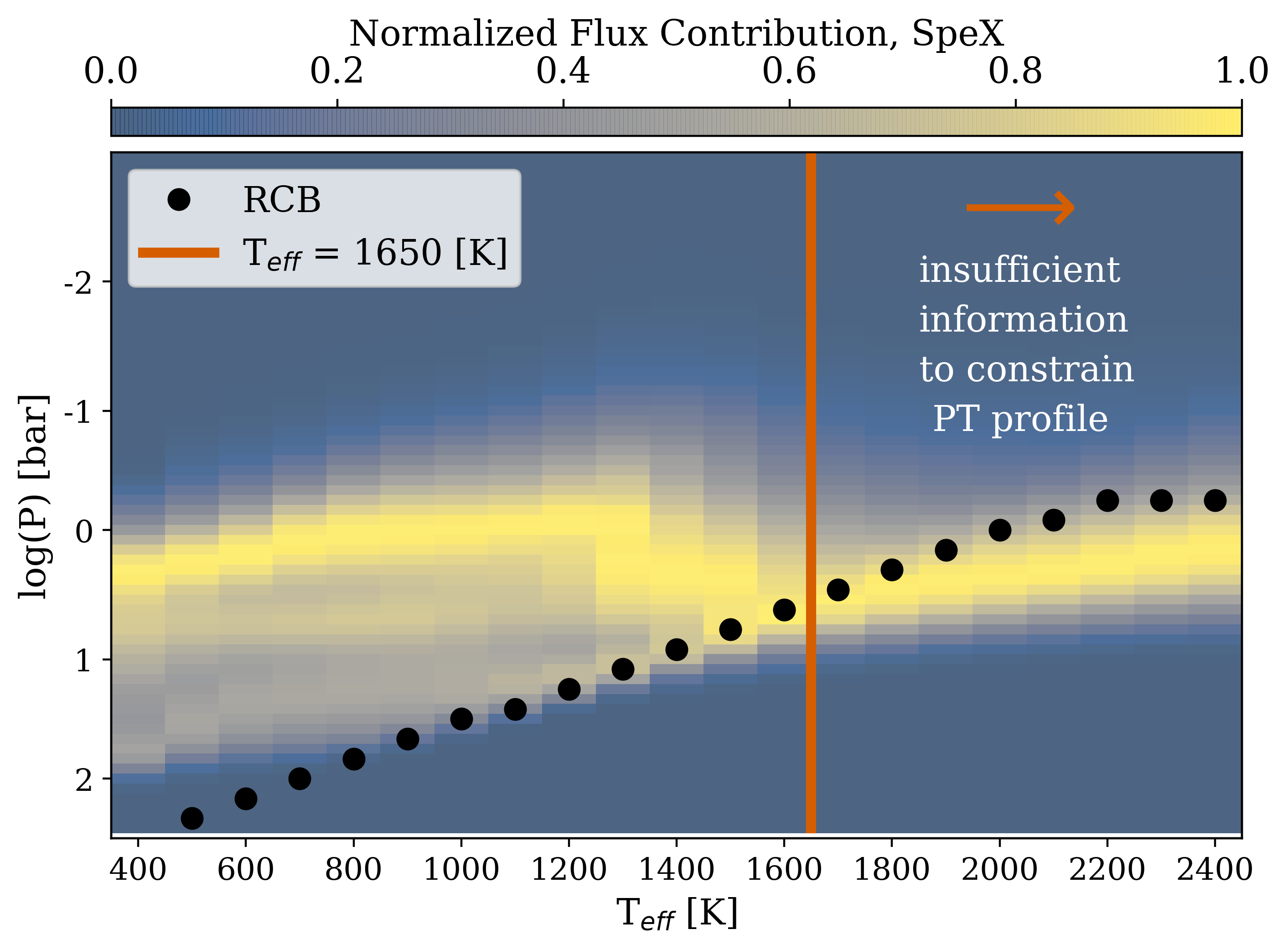}
    \includegraphics[width=0.47\linewidth]{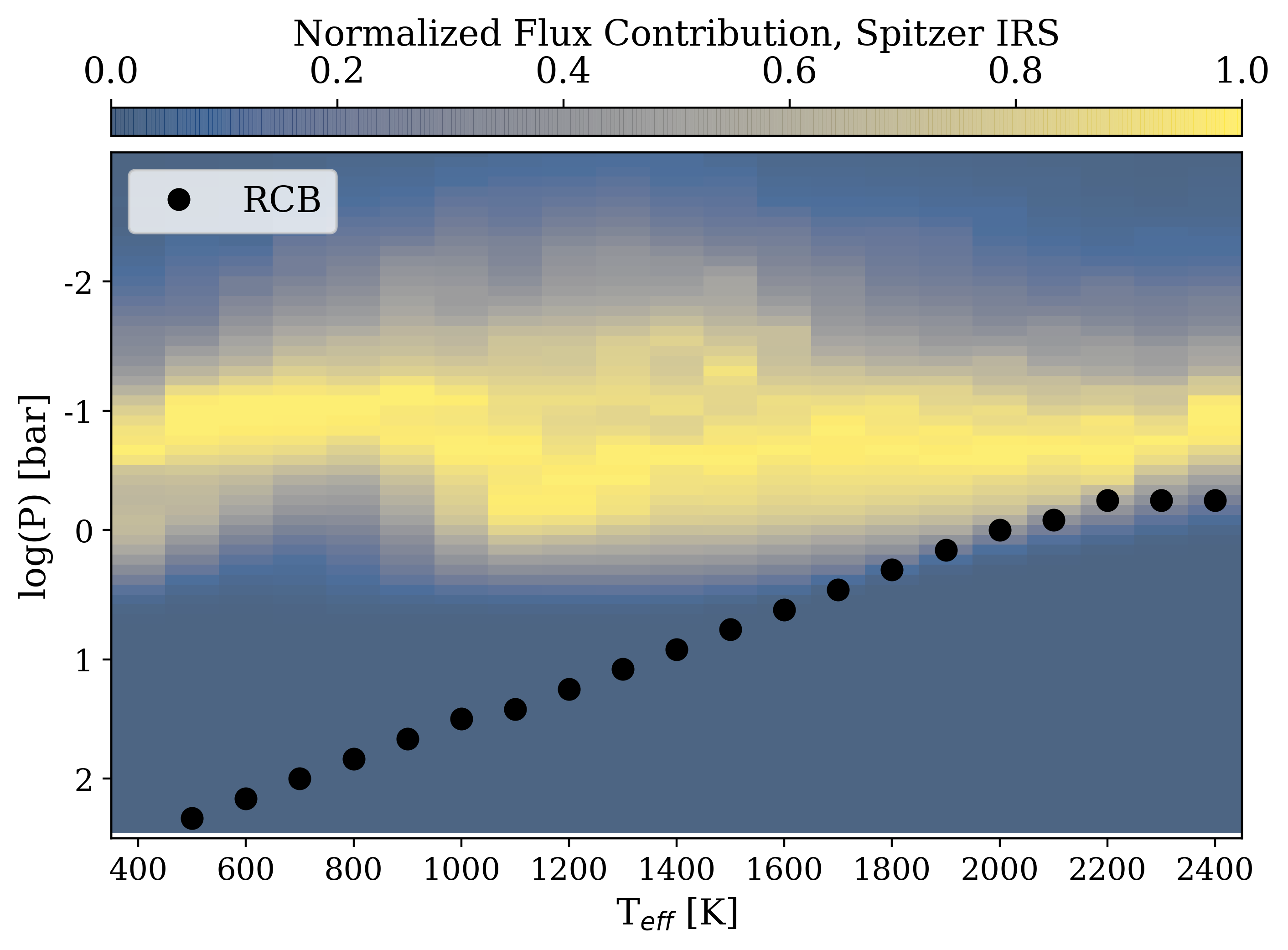}
    \caption{The RCB from cloud-free Sonora models for objects with T$_{\rm eff}$ = 500 to 2400 K. Shading indicates SpeX ($\lambda$ = 1.0 - 2.5 $\um$, R $\approx$ 130) [left] normalized contributed flux from each layer. Fewer layers are probed at hotter temperatures, which makes thermal profile characterization difficult for NIR spectra. The same plot for Spitzer IRS $\lambda$ = 5.2 - 14 $\um$, R $\approx$ 90) coverage [right] showing additional layers probed with extended wavelength coverage.}
    \label{fig:rcb_phot}
\end{figure*}

We find that the ``smoothed free'' PT profile prescriptions used in prior brown dwarf retrieval models \citep{line2015} do not successfully reproduce the upper atmospheres of early-L dwarfs: the parameters that prioritize smoothness end up creating a linear PT profile that is cooler than the true profile for objects above $\sim$1700 K. Here we explain the physical reason for this, and provide a rule-of-thumb for future L dwarf retrievals. 

Thermal profile characterization of these hotter objects is difficult because of the number and location of pressure layers probed with NIR spectra. Radiative-convective profiles have a characteristic change in slope at the boundary between the convective layers (which are steeper) and the radiative layers (which are more isothermal). This radiative-convective boundary (RCB) rises to higher altitudes for hotter objects; for objects warmer than about 1700 K, NIR spectra probe mostly the convective regions of the atmosphere. Because they do not adequately capture the curvature at the RCB, all PT profile parameterizations will struggle to reproduce that structure robustly. 

Figure \ref{fig:rcb_phot}  shows the RCB from the cloud-free Sonora models with T$_{\rm eff}$ = 500 to 2400 K and shading indicating the normalized contributed flux at each layer from SpeX ($\lambda$ = 1.0 - 2.5 $\um$, R $\approx$ 130) spectra. Fewer layers are probed at hotter temperatures even in the absence of clouds. The orange line in Figure \ref{fig:rcb_phot} indicates where the smoothed PT profile parameterization started to retrieve cooler than expected upper atmospheres (similarly, an unsmoothed PT profile was unconstrained outside of the probed pressure layers, though still captured the true PT profile within the 1$\sigma$ uncertainties). The narrow range of pressures probed in the L2 spectrum also impacted more structured PT profiles like the grey and nongrey RC profiles, which retrieved hotter profiles deep in the atmosphere and resulted in incorrect retrieved abundances, metallicities, and surface gravities.

The right plot of Figure \ref{fig:rcb_phot} shows the same information but for Spitzer IRS ($\lambda$ = 5.2 - 14 $\um$, R $\approx$ 90) coverage. The mid-infrared spectra probe lower pressures since the opacities of species generally increase from near- to mid-infrared wavelengths; the pressures probed are located in the radiative part of the atmosphere. We therefore find that combining near- and mid-infrared data of early-mid L dwarfs probes convective regions (in the near-infrared) and radiative regions (in the mid-infrared), giving the retrieval information about the PT profile curvature. 

This agrees with previous retrievals of 2M2224-0158, a cloudy L4.5 dwarf. \citet{burningham2017} were unable to retrieve a PT profile with SpeX NIR data alone, but \citet{burningham2021} were able to constrain both the PT profile and some cloud properties when combining NIR and MIR SpeX, IRCS L' band, and Spitzer IRS data for the object. While we successfully constrained bulk properties like surface gravity, effective temperature, metallicity, and C/O ratios with only NIR data, MIR data is needed to robustly determine the thermal profile and cloud properties for early- to mid-L dwarfs.

Our rule-of-thumb for future retrievals is as follows: for an object with the estimated T$_{\rm{\rm eff}}$/log g of your target, use a self-consistent model to understand whether your observations probe both above and below the radiative-convective boundary. If observations only probe the deep convective atmosphere, proceed with caution with parameterized PT profiles and do not trust the retrieved thermal profile above the RC boundary. 

\subsection{``Free'' Chemistry Approaches in Brown Dwarfs}
When conducting retrieval tests on the cloud-free L2 self-consistent chemistry spectrum, ignoring the effects of rainout chemistry and strongly nonuniform abundances resulted in skewed surface gravities, metallicities, C/O ratios, and incorrect PT profiles. 

We determined that a nonuniform (step function) FeH profile resulted in accurate bulk properties and photospheric thermal profiles. This best fit model retrieved uniform abundances for all species except FeH and was strongly preferred over all other free chemistry parameterizations.

Nonuniform FeH chemistry is needed to accurately retrieve abundances in this object because FeH is both a strong absorber in the NIR and is heavily affected by rainout chemistry. In this section, we expand the analysis to predict at what temperatures other species are likely to have both strongly nonuniform abundances and a significant impact on NIR spectra; these are the species most likely to warrant including nonuniform abundances in retrievals.

\subsubsection{Applicability of Nonuniform Chemistry}
\begin{figure*}[!htb]
    \centering
    \includegraphics[width=0.465\linewidth]{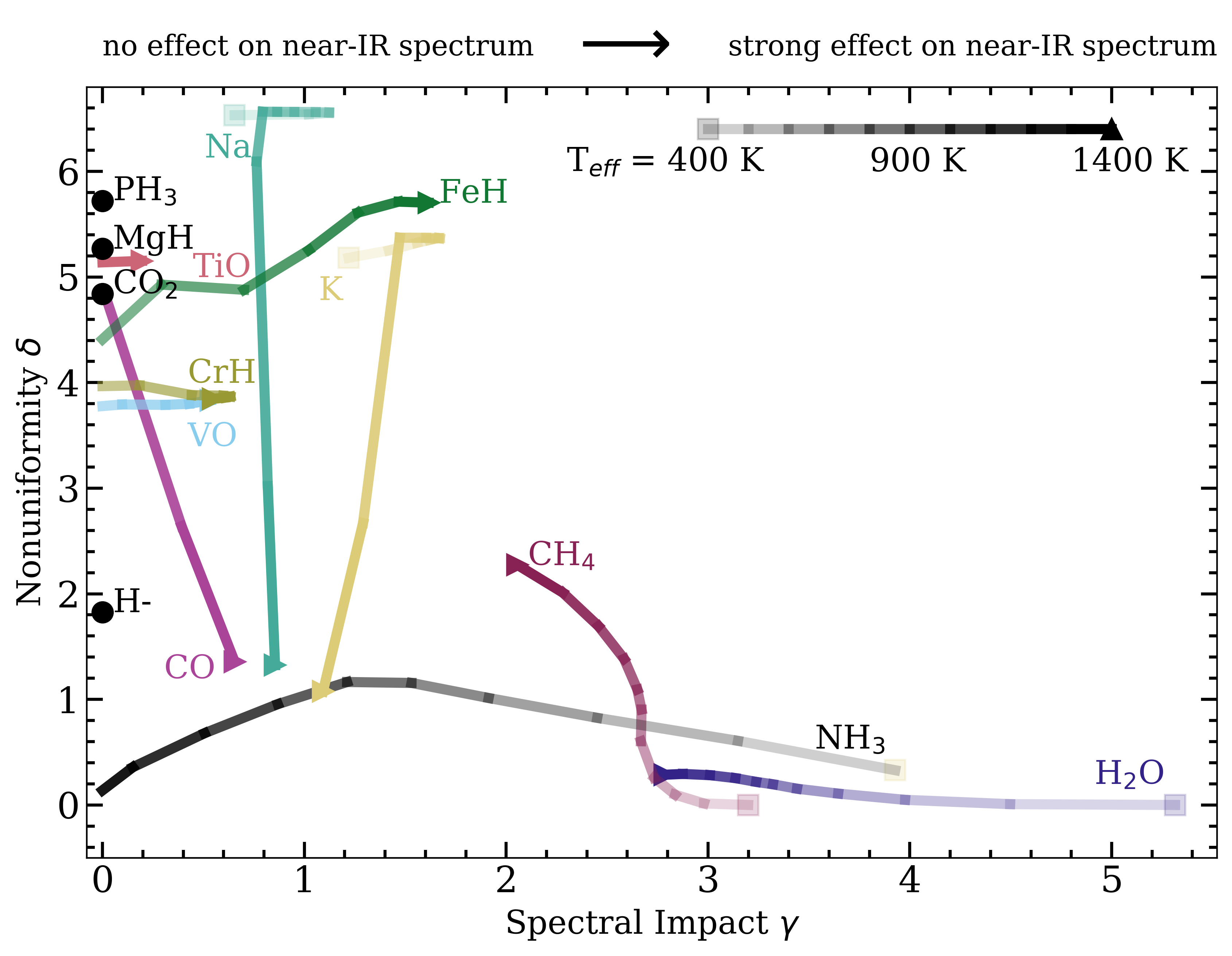}
    \includegraphics[width=0.495\linewidth]{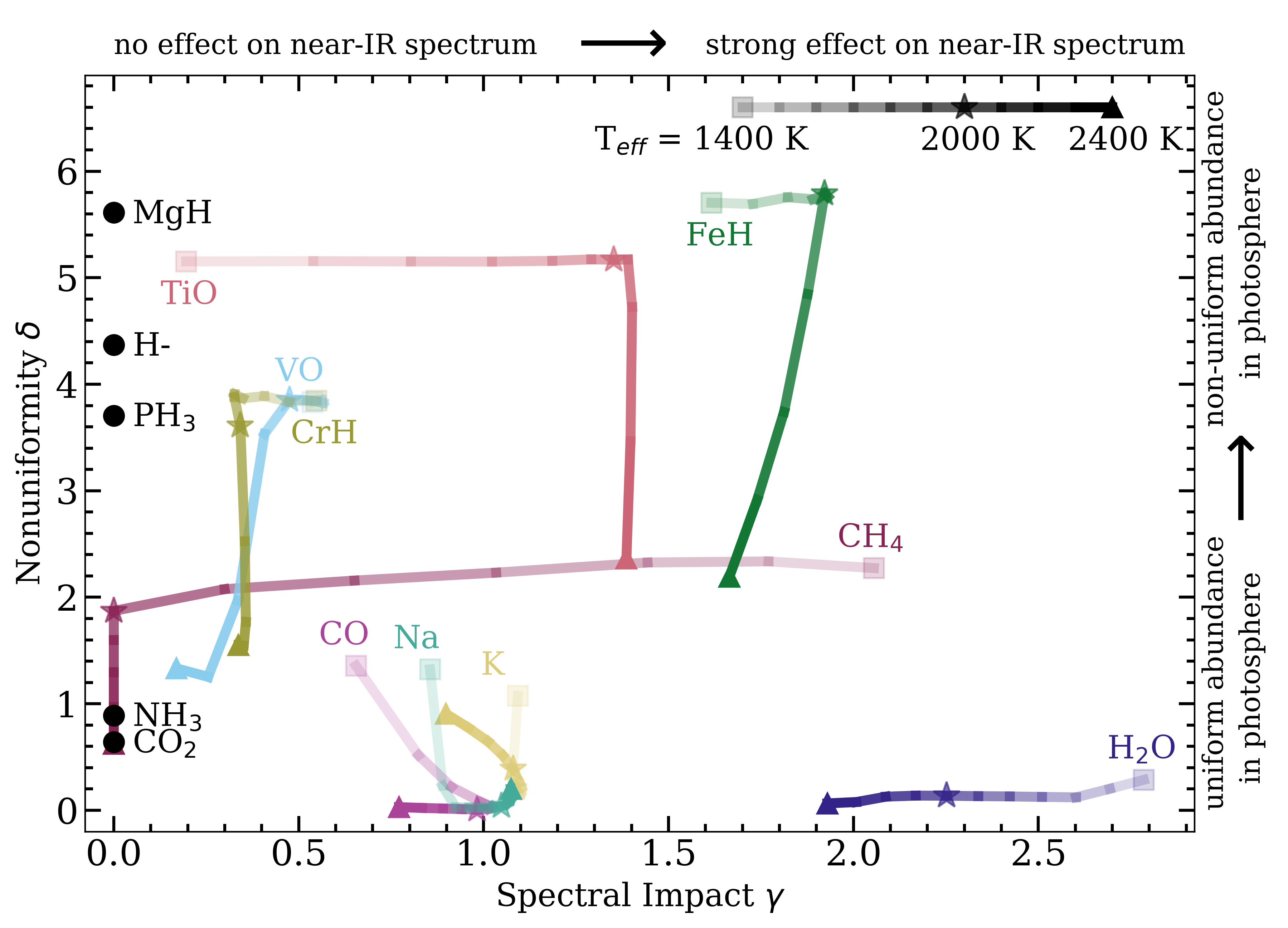}
    \caption{Nonuniformity vs spectral impact of 14 gases for T and Y dwarfs [left] and L dwarfs [right]. Color gradients indicate different effective temperatures. Gases in the upper right of each plot have high spectral impact and have highly nonuniform abundance profiles. Gases whose removal does not impact any data points in the spectrum above the noise are indicated with black circles. The stars on the right plot indicate the values for the 2000 K object studied in this paper.}
    \label{fig:gassummary}
\end{figure*}

A more complex abundance profile for a given gas would only be warranted if the gas both strongly impacts the observed spectrum and is expected to be strongly nonuniform in the atmosphere.  We defined both a ``nonuniformity index'' and a ``spectral impact index'' to determine which gases may warrant a more complex abundance profile in objects with a range of effective temperatures. 

Nonuniformity $\delta$ was determined by the change in mixing ratio of the gas in the photosphere using equation \ref{eq:nonuniformity}. 

\begin{equation}
\delta = \max(\log_{10}f_{i}) - \min(\log_{10}f_{i})
\label{eq:nonuniformity}
\end{equation}
where $f_{i}$ is the VMR of a gas abundance profile at level \textit{i} in the atmosphere. We restrict the atmospheric layers considered to the photosphere, which is determined by the contributed flux in each layer of the atmosphere. We impose a lower limit of min(log$_{10}f_{i}$) = -12. 

The spectral impact $\gamma$ is determined by comparing the Sonora spectrum for the object to the spectrum with the gas removed using equation \ref{eq:spectralimpact}. 

\begin{equation}
\gamma = \log_{10}\max(\frac{F_{gas,i} - F_{Sonora,i}}{F_{Sonora,i}}) + 1.52
\label{eq:spectralimpact}
\end{equation}
    where $F_{gas,i}$ is the flux from the spectrum with a gas removed at wavelength \textit{i} and $F_{Sonora,i}$ is the flux from the Sonora spectrum at wavelength \textit{i} for SpeX data (R = 120, $\lambda$ = 1.0 - 2.5 $\um$). The values were shifted by 1.52 so that a value of 0 aligns with a spectral impact less than the mean spectral noise. Figure \ref{fig:Sonora_gas} in Appendix \ref{AppendixB} shows the effect of varying the abundance of each gas for a T$_{\rm eff}$ = 2000 K spectrum and can be used to inform gases included in future retrievals. Small changes ($\pm$0.3 dex) in gases like H$_{2}$O and FeH have large impacts on the spectrum while large changes ($\pm$2.0 dex) in gases like CO$_{2}$ and H- have no effect on the spectrum.

We used the Sonora spectra at NIR SpeX resolution and the Sonora abundance profiles to calculate these indices for the 14 gas species in this paper in 21 objects with T$_{\rm eff}$ ranging from 400 K to 2400 K (log(g) = 5.0, solar metallicity and C/O), and the results are displayed in Figure \ref{fig:gassummary}. The gases for the object studied in this paper (T$_{\rm eff}$ = 2000 K) are indicated by stars in the right plot. 

A nonuniform FeH abundance profile was needed to accurately retrieve bulk properties for a 2000 K object. Figure \ref{fig:gassummary} shows that FeH has high nonuniformity and high spectral impact between 1800 and 2200 K, but becomes more uniform at hotter temperatures as rainout occurs above the photosphere. This holds true for all species with prominent rainout features above 2200 K. \citet{zalesky2019} found that alkalis are affected by rainout in late-T and early-Y spectra. Similarly, we find that alkalis have high nonuninformity and relatively high spectral impact in early-Y spectra (T$_{\rm eff}$ $\approx$ 600 K). A simpler 2-parameter nonuniform profile (deep atmosphere abundance and cut-off pressure) might be considered for species that are expected to undergo total depletion via rainout. Additionally, CH$_{4}$ has high spectral impact at starting with late-L dwarfs. While not affected by rainout, its abundance profile can vary by more than two orders of magnitude in the photosphere. Therefore, these gases may warrant nonuniform abundance profiles in future retrieval work.

Figure \ref{fig:gassummary} is only applicable for low resolution NIR data, and would change depending on wavelength coverage and resolution. Deviations from chemical equilibrium and solar abundances would also change values. Oxide (particularly TiO) abundances are expected to have strongly nonuniform profiles due to rainout near the photosphere in objects as cool as 1400 K, however their features are most strongly seen in the optical.

Our rule-of-thumb for future retrievals is that practitioners should use self-consistent models to guide their analysis: any spectrally active gases likely to vary by more than 2 orders of magnitude in abundance within the photosphere should warrant additional tests to see if a nonuniform abundance should be used. 

\section{Conclusions}\label{Conclusions}
Characterizing L dwarfs is challenging due to their narrow NIR photospheres, the rainout of prominent opacity sources, and the presence of clouds. This work sought to inform future retrievals by addressing the first two of these challenges. 
We conducted atmospheric retrievals on synthetic cloud-free L dwarf SpeX spectra derived from the Sonora Bobcat models.  We tested a variety of PT profile and abundance profile parameterizations to determine how they bias retrieved bulk properties such as surface gravity, effective temperature, metallicity, and C/O ratios. 

\begin{enumerate}
    \item For early-L dwarfs, parameterized PT profiles retrieved biased results for most bulk properties. Free, unsmoothed PT profiles accurately retrieved all bulk properties.   
    \item Both NIR and MIR data are needed to constrain PT profiles in early- to mid-L dwarfs because NIR data alone probes mostly convective regions of their atmospheres, below the radiative-convective boundary. 
    \item For atmospheres with nonuniform species,  assuming that all gases have uniform abundances causes the retrieved gravity, metallicity, and C/O ratios to be incorrect, regardless of PT profile parameterization. A nonuniform (step function) abundance profile for FeH was needed to accurately retrieve bulk properties for an L2 dwarf. 
    \item Nonuniform FeH is needed for early- to mid-L NIR retrievals. Other rainout species like TiO may need nonuniform treatment at optical wavelengths. Nonuniform prescriptions may also be important near the L/T transition (CH$_{4}$) and early-Y dwarfs (Na and K) with the same spectral coverage. 
\end{enumerate}

We have demonstrated the utility of using self-consistent models to guide retrievals. We used sophisticated self-consistent models to assess the validity of retrieval techniques for L dwarfs, finding several shortcomings of prior approaches. We presented two rules-of-thumb for practitioners.  First, consider the location of the radiative-convective boundary compared to the pressures probed by your observations; aim to collect data that probes both radiative and convective layers to constrain the key curvature of the profile, and be cautious applying parameterized PT profiles without this. Second, consider spectrally-active gases likely to vary by more than two orders of magnitude in abundance in the photosphere, and include tests to assess nonuniformity. Future retrievals should also consider doing tests such as those presented here using self-consistent models rather than toy models, to be sure that their approach is unbiased and can capture the complexity of real substellar atmospheres.

\section{Acknowledgements}
The authors acknowledge the Texas Advanced Computing Center (TACC) at The University of Texas at Austin for providing high performance computing resources that have contributed to the research results reported within this paper. URL: http://www.tacc.utexas.edu. MJR acknowledges funding from NASA FINESST grant NNH19ZDA001N-FINESST. CVM acknowledges funding from National Science Foundation AAG grant 1910969.

\textit{Software}: CHIMERA \citep{line2013b}, \texttt{emcee} \citep{foremanmackey2013}, \texttt{corner.py} \citep{foremanmackey2016}

\appendix
\section{Retrieved Abundances}\label{AppendixAA}
\begin{figure*}[!hb]
    \centering
    \includegraphics[width=0.83\textwidth]{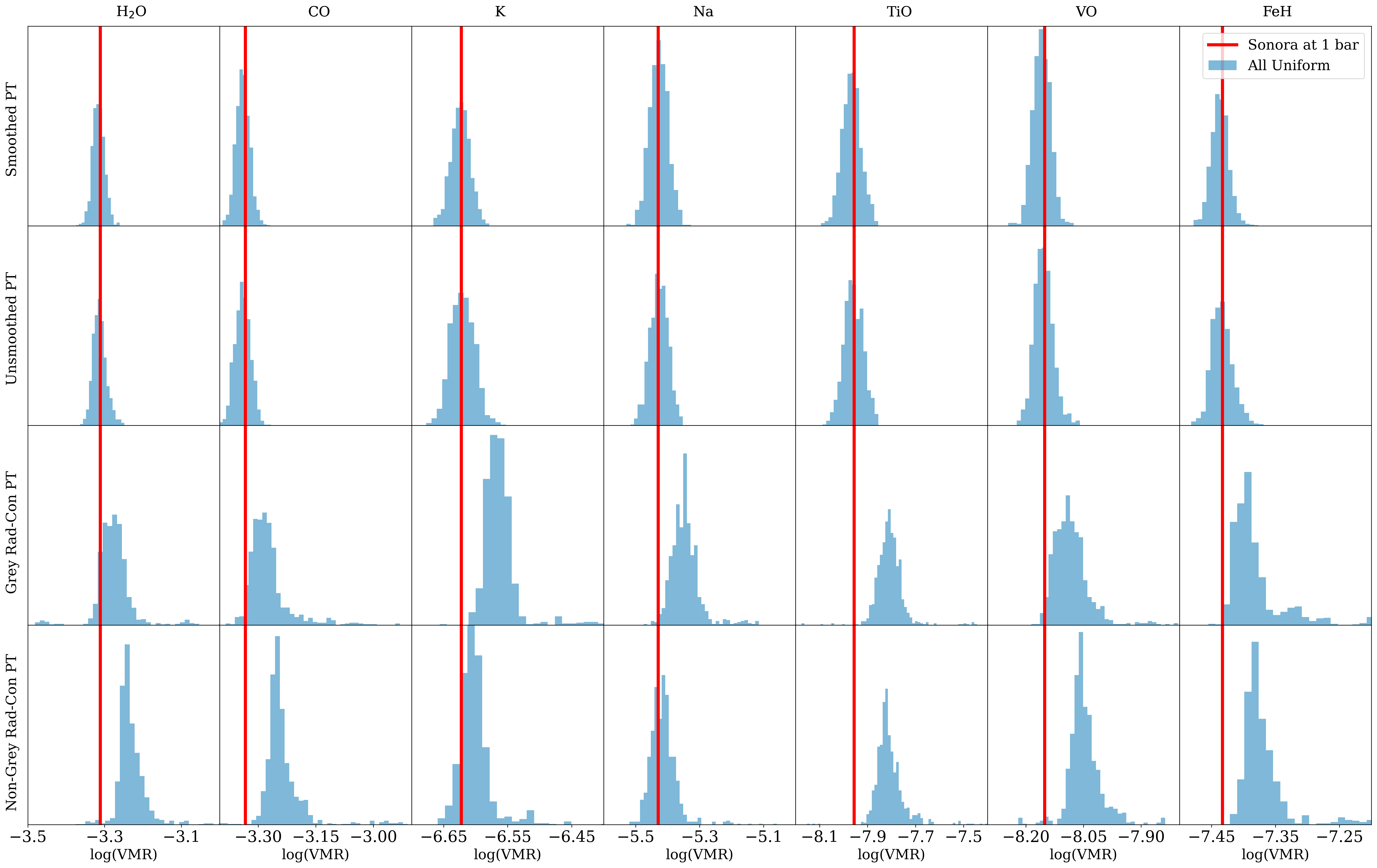}
    \includegraphics[width=0.83\textwidth]{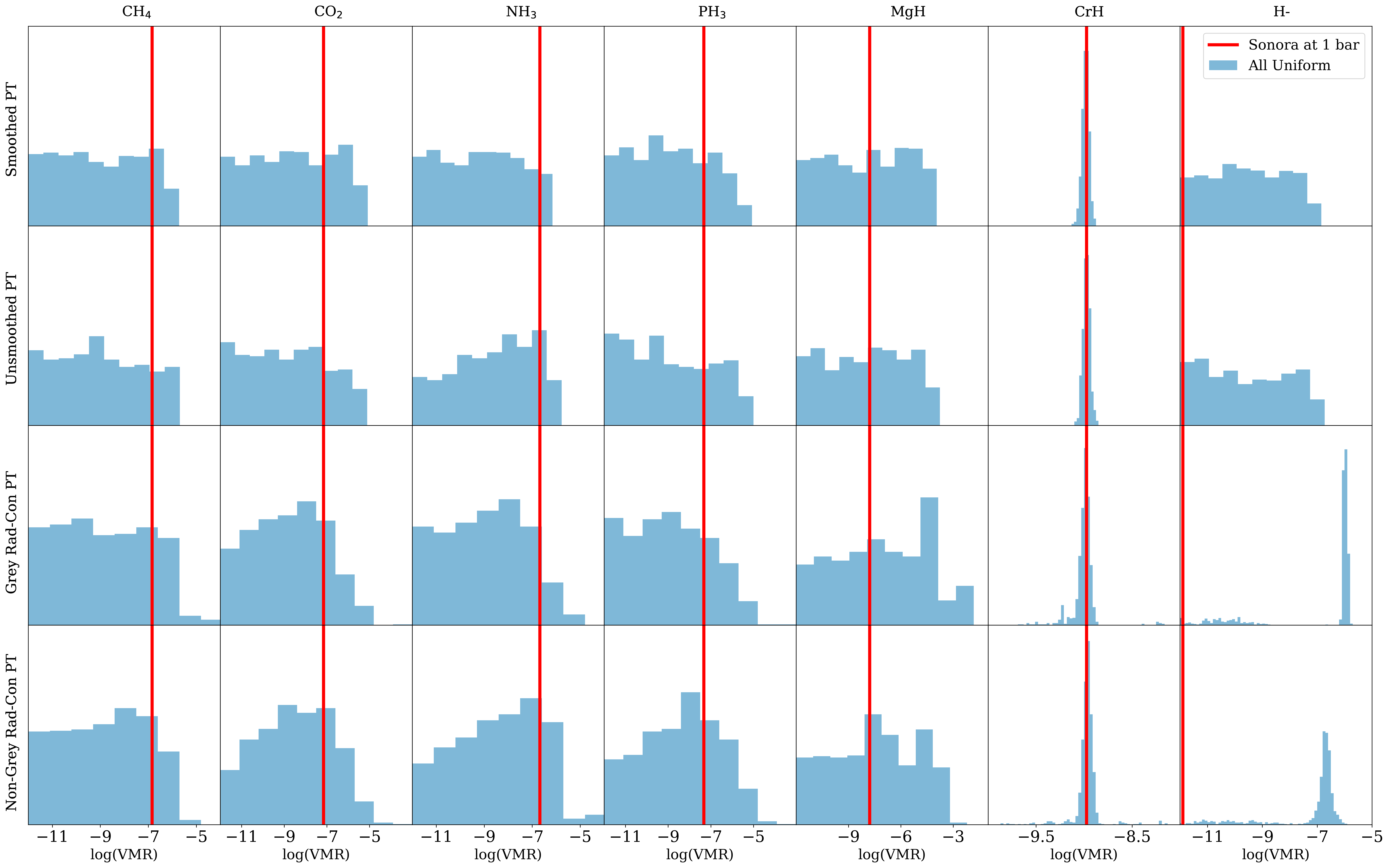}
    \caption{The posteriors of the uniform abundance parameters [blue] and the input uniform chemistry profiles [red] for the 4 thermal profiles tested. The top plot shows constrained species and the bottom plot shows mostly unconstrained species.}
    \label{fig:uniabund_append}
\end{figure*}
\begin{figure*}[!ht]
    \centering
    \includegraphics[width=0.85\textwidth]{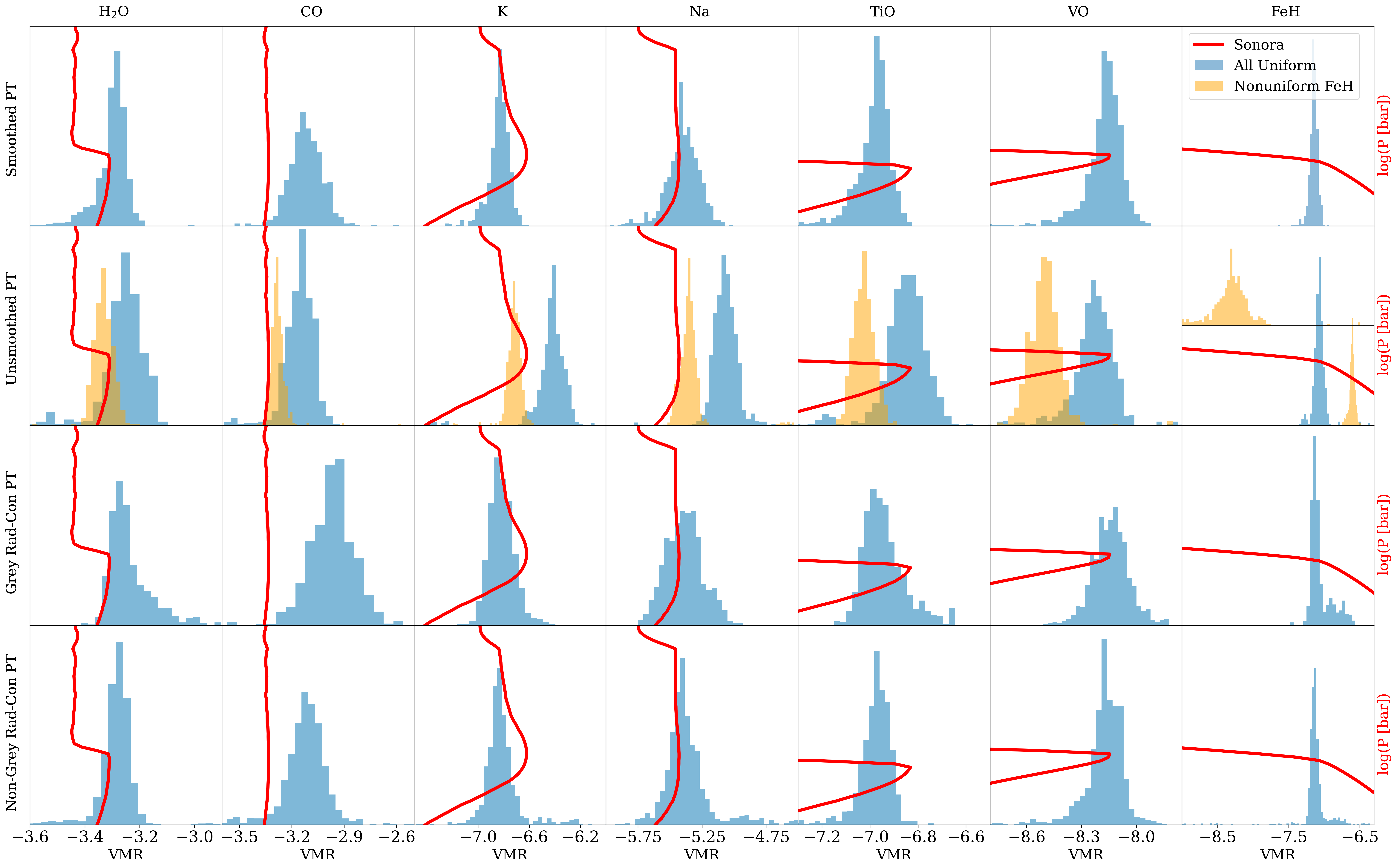}
    \includegraphics[width=0.85\textwidth]{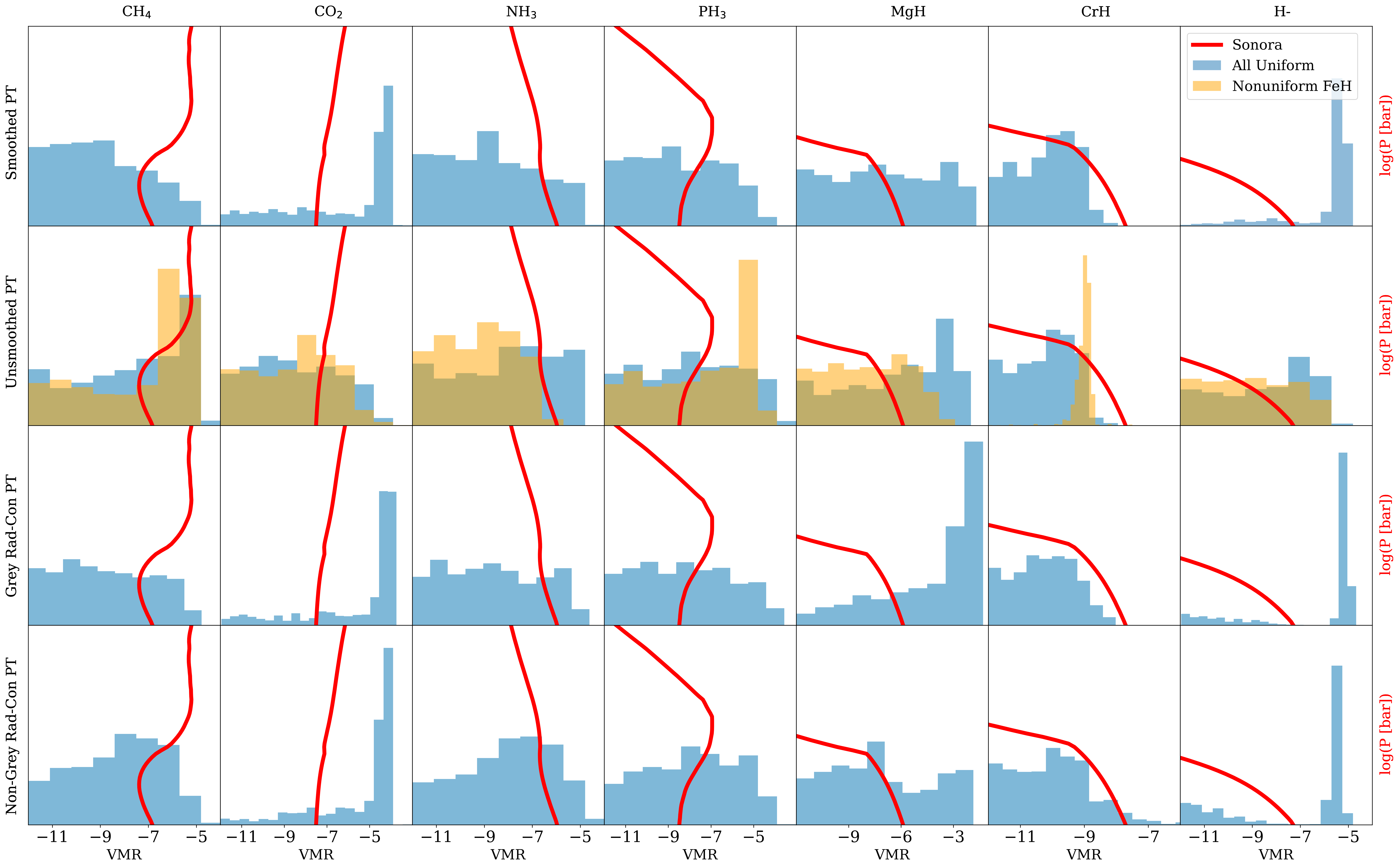}
    \caption{The posteriors of the fully uniform abundance parameters [blue] and the input self-consistent chemistry abundance profiles [red] for the 4 thermal profiles tested. Also shown are the posteriors for the non-uniform FeH retrieval [orange]. The top plot shows constrained species and the bottom plot shows mostly unconstrained species. Both the upper and lower abundance for the non-uniform FeH posterior are shown.}
    \label{fig:nonuniabund_append}
\end{figure*}
\clearpage
\section{Nonuniform FeH Retrieval}\label{AppendixA}
\begin{figure*}[!ht]
    \centering
    \includegraphics[width=1.0\textwidth]{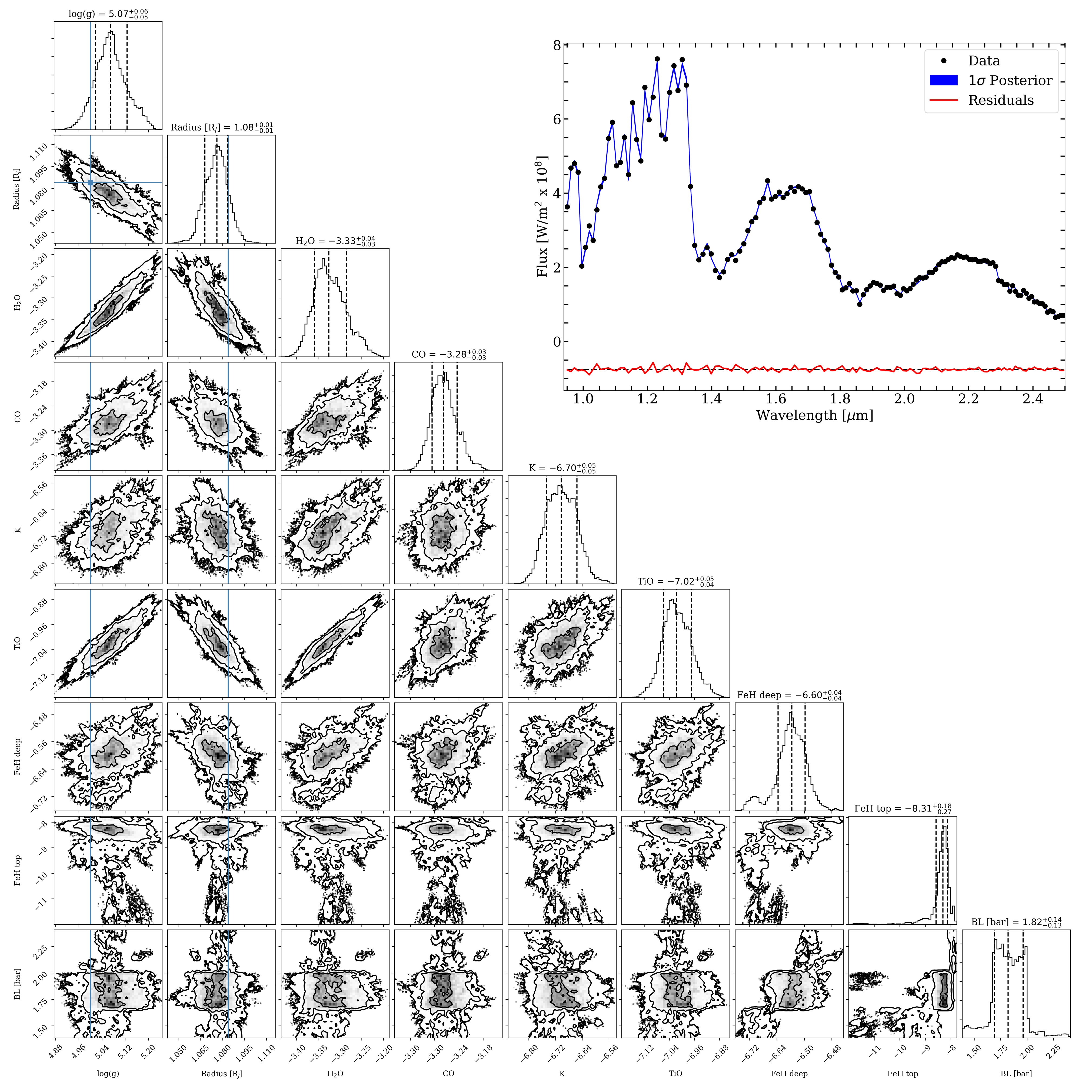}
    \caption{Corner plot of selected retrieved parameters including the 6 most abundant species for the nonuniform FeH retrieval. All abundances are log(VMR) values. The true values for surface gravity are indicated in blue. Upper figure is the 1$\sigma$ retrieved spectra of the nonuniform FeH model with residuals.}
    \label{fig:fehretrieval}
\end{figure*}
\clearpage
\section{Effects of Gases on L2 NIR SpeX Spectra}\label{AppendixB}
\begin{figure*}[!ht]
    \centering
    \includegraphics[width=1.0\textwidth]{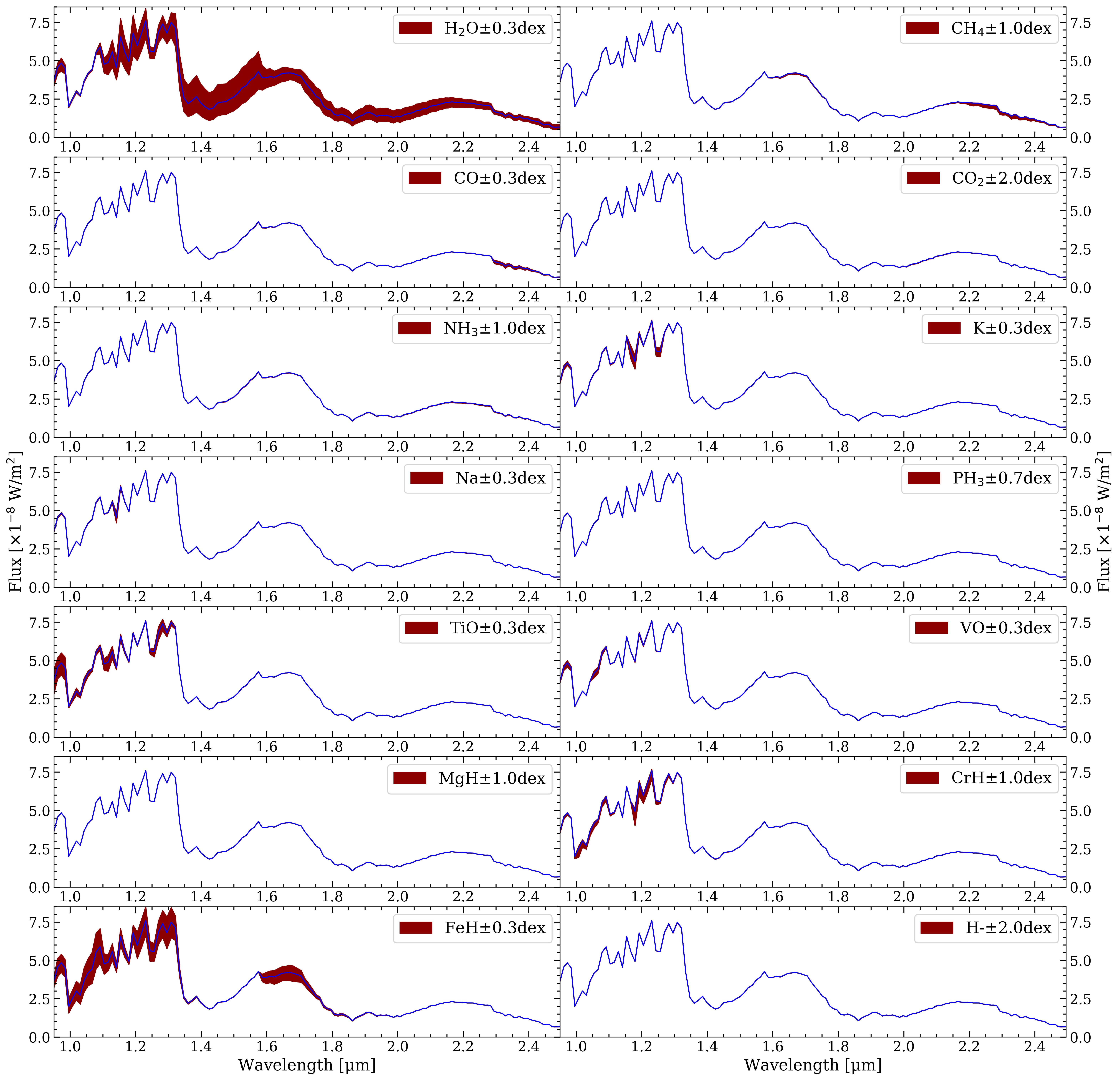}
    \caption{CHIMERA forward models using the Sonora PT profile and abundance profiles [blue] and the effect of changing each gas abundance profile $\pm$ X dex according to allowable amount based on solar maximum. A dex of $\pm$1.0 changes the abundance by 10 $\times$ the Sonora value and shifts the abundance profile by 1 in log(VMR) space. The resulting spectra are not self-consistent.}
    \label{fig:Sonora_gas}
\end{figure*}
\clearpage
\bibliographystyle{aasjournal}
\bibliography{references}

\begin{thebibliography}{}
\expandafter\ifx\csname natexlab\endcsname\relax\def\natexlab#1{#1}\fi
\providecommand{\url}[1]{\href{#1}{#1}}

\bibitem[{{Ackerman} \& {Marley}(2001)}]{ackermanmarley2001}
{Ackerman}, A.~S., \& {Marley}, M.~S. 2001, \apj, 556, 872

\bibitem[{{Allard} {et~al.}(2000){Allard}, {Hauschildt}, \&
  {Schwenke}}]{Allard2000}
{Allard}, F., {Hauschildt}, P.~H., \& {Schwenke}, D. 2000, \apj, 540, 1005

\bibitem[{{Barber} {et~al.}(2006){Barber}, {Tennyson}, {Harris}, \&
  {Tolchenov}}]{Barber2006}
{Barber}, R.~J., {Tennyson}, J., {Harris}, G.~J., \& {Tolchenov}, R.~N. 2006,
  \mnras, 368, 1087

\bibitem[{{Bourrier} {et~al.}(2020){Bourrier}, {Kitzmann}, {Kuntzer},
  {Nascimbeni}, {Lendl}, {Lavie}, {Hoeijmakers}, {Pino}, {Ehrenreich}, {Heng},
  {Allart}, {Cegla}, {Dumusque}, {Melo}, {Astudillo-Defru}, {Caldwell},
  {Cretignier}, {Giles}, {Henze}, {Jenkins}, {Lovis}, {Murgas}, {Pepe},
  {Ricker}, {Rose}, {Seager}, {Segransan}, {Su{\'a}rez-Mascare{\~n}o}, {Udry},
  {Vanderspek}, \& {Wyttenbach}}]{Bourrier2020}
{Bourrier}, V., {Kitzmann}, D., {Kuntzer}, T., {et~al.} 2020, \aap, 637, A36

\bibitem[{{Burgasser}(2014)}]{burgasser2014}
{Burgasser}, A.~J. 2014, in Astronomical Society of India Conference Series,
  Vol.~11, Astronomical Society of India Conference Series, 7--16

\bibitem[{{Burningham} {et~al.}(2017){Burningham}, {Marley}, {Line}, {Lupu},
  {Visscher}, {Morley}, {Saumon}, \& {Freedman}}]{burningham2017}
{Burningham}, B., {Marley}, M.~S., {Line}, M.~R., {et~al.} 2017, \mnras, 470,
  1177

\bibitem[{{Burningham} {et~al.}(2021){Burningham}, {Faherty}, {Gonzales},
  {Marley}, {Visscher}, {Lupu}, {Gaarn}, {Fabienne Bieger}, {Freedman}, \&
  {Saumon}}]{burningham2021}
{Burningham}, B., {Faherty}, J.~K., {Gonzales}, E.~C., {et~al.} 2021, \mnras,
  506, 1944

\bibitem[{{Burrows} {et~al.}(2001){Burrows}, {Hubbard}, {Lunine}, \&
  {Liebert}}]{burrows2001}
{Burrows}, A., {Hubbard}, W.~B., {Lunine}, J.~I., \& {Liebert}, J. 2001,
  Reviews of Modern Physics, 73, 719

\bibitem[{{Burrows} {et~al.}(1993){Burrows}, {Hubbard}, {Saumon}, \&
  {Lunine}}]{burrows1993}
{Burrows}, A., {Hubbard}, W.~B., {Saumon}, D., \& {Lunine}, J.~I. 1993, \apj,
  406, 158

\bibitem[{{Burrows} {et~al.}(2002){Burrows}, {Ram}, {Bernath}, {Sharp}, \&
  {Milsom}}]{burrows2002}
{Burrows}, A., {Ram}, R.~S., {Bernath}, P., {Sharp}, C.~M., \& {Milsom}, J.~A.
  2002, \apj, 577, 986

\bibitem[{{Chandrasekhar}(1935)}]{chandrasekhar1935}
{Chandrasekhar}, S. 1935, \mnras, 96, 21

\bibitem[{{Changeat} {et~al.}(2019){Changeat}, {Edwards}, {Waldmann}, \&
  {Tinetti}}]{changeat2019}
{Changeat}, Q., {Edwards}, B., {Waldmann}, I.~P., \& {Tinetti}, G. 2019, \apj,
  886, 39

\bibitem[{{Cushing} {et~al.}(2008){Cushing}, {Marley}, {Saumon}, {Kelly},
  {Vacca}, {Rayner}, {Freedman}, {Lodders}, \& {Roellig}}]{cushing2008}
{Cushing}, M.~C., {Marley}, M.~S., {Saumon}, D., {et~al.} 2008, \apj, 678, 1372

\bibitem[{{Dulick} {et~al.}(2003){Dulick}, {Bauschlicher}, {Burrows}, {Sharp},
  {Ram}, \& {Bernath}}]{dulick2003}
{Dulick}, M., {Bauschlicher}, C.~W., J., {Burrows}, A., {et~al.} 2003, \apj,
  594, 651

\bibitem[{{Foreman-Mackey}(2016)}]{foremanmackey2016}
{Foreman-Mackey}, D. 2016, The Journal of Open Source Software, 1, 24

\bibitem[{{Foreman-Mackey} {et~al.}(2013){Foreman-Mackey}, {Hogg}, {Lang}, \&
  {Goodman}}]{foremanmackey2013}
{Foreman-Mackey}, D., {Hogg}, D.~W., {Lang}, D., \& {Goodman}, J. 2013, \pasp,
  125, 306

\bibitem[{{Gonzales} {et~al.}(2022){Gonzales}, {Burningham}, {Faherty},
  {Lewis}, {Visscher}, \& {Marley}}]{gonzales2022arXiv220902754G}
{Gonzales}, E.~C., {Burningham}, B., {Faherty}, J.~K., {et~al.} 2022, arXiv
  e-prints, arXiv:2209.02754

\bibitem[{{Gonzales} {et~al.}(2021){Gonzales}, {Burningham}, {Faherty},
  {Visscher}, {Marley}, {Lupu}, {Freedman}, \& {Lewis}}]{gonzales2021L3}
---. 2021, \apj, 923, 19

\bibitem[{{Gravity Collaboration} {et~al.}(2020){Gravity Collaboration},
  {Nowak}, {Lacour}, {Molli{\`e}re}, {Wang}, {Charnay}, {van Dishoeck},
  {Abuter}, {Amorim}, {Berger}, {Beust}, {Bonnefoy}, {Bonnet}, {Brandner},
  {Buron}, {Cantalloube}, {Collin}, {Chapron}, {Cl{\'e}net}, {Coud{\'e} Du
  Foresto}, {de Zeeuw}, {Dembet}, {Dexter}, {Duvert}, {Eckart}, {Eisenhauer},
  {F{\"o}rster Schreiber}, {F{\'e}dou}, {Garcia Lopez}, {Gao}, {Gendron},
  {Genzel}, {Gillessen}, {Hau{\ss}mann}, {Henning}, {Hippler}, {Hubert},
  {Jocou}, {Kervella}, {Lagrange}, {Lapeyr{\`e}re}, {Le Bouquin}, {L{\'e}na},
  {Maire}, {Ott}, {Paumard}, {Paladini}, {Perraut}, {Perrin}, {Pueyo}, {Pfuhl},
  {Rabien}, {Rau}, {Rodr{\'\i}guez-Coira}, {Rousset}, {Scheithauer},
  {Shangguan}, {Straub}, {Straubmeier}, {Sturm}, {Tacconi}, {Vincent},
  {Widmann}, {Wieprecht}, {Wiezorrek}, {Woillez}, {Yazici}, \&
  {Ziegler}}]{gravity2020}
{Gravity Collaboration}, {Nowak}, M., {Lacour}, S., {et~al.} 2020, \aap, 633,
  A110

\bibitem[{{Guillot}(2010)}]{guillot2010}
{Guillot}, T. 2010, \aap, 520, A27

\bibitem[{{Hargreaves} {et~al.}(2010){Hargreaves}, {Hinkle}, {Bauschlicher},
  {Wende}, {Seifahrt}, \& {Bernath}}]{hargreaves2010}
{Hargreaves}, R.~J., {Hinkle}, K.~H., {Bauschlicher}, Charles~W., J., {et~al.}
  2010, \aj, 140, 919

\bibitem[{{Huang} {et~al.}(2014){Huang}, {Gamache}, {Freedman}, {Schwenke}, \&
  {Lee}}]{Huang2014}
{Huang}, X., {Gamache}, R.~R., {Freedman}, R.~S., {Schwenke}, D.~W., \& {Lee},
  T.~J. 2014, \jqsrt, 147, 134

\bibitem[{{Kass} \& {Raftery}(1995)}]{kassraftery1995}
{Kass}, R.~E., \& {Raftery}, A.~E. 1995, Journal of the American Statistical
  Association, 90, 773

\bibitem[{{Kitzmann} {et~al.}(2020){Kitzmann}, {Heng}, {Oreshenko}, {Grimm},
  {Apai}, {Bowler}, {Burgasser}, \& {Marley}}]{kitzmann2020heliosr2}
{Kitzmann}, D., {Heng}, K., {Oreshenko}, M., {et~al.} 2020, \apj, 890, 174

\bibitem[{{Lacis} \& {Oinas}(1991)}]{lacisoinas1991}
{Lacis}, A.~A., \& {Oinas}, V. 1991, \jgr, 96, 9027

\bibitem[{{Lacy}(submitted)}]{lacy2022}
{Lacy}, B. submitted, \apj

\bibitem[{{Li} {et~al.}(2015){Li}, {Gordon}, {Rothman}, {Tan}, {Hu}, {Kassi},
  {Campargue}, \& {Medvedev}}]{li2015}
{Li}, G., {Gordon}, I.~E., {Rothman}, L.~S., {et~al.} 2015, \apjs, 216, 15

\bibitem[{{Line} {et~al.}(2015){Line}, {Teske}, {Burningham}, {Fortney}, \&
  {Marley}}]{line2015}
{Line}, M.~R., {Teske}, J., {Burningham}, B., {Fortney}, J.~J., \& {Marley},
  M.~S. 2015, \apj, 807, 183

\bibitem[{{Line} {et~al.}(2013){Line}, {Wolf}, {Zhang}, {Knutson}, {Kammer},
  {Ellison}, {Deroo}, {Crisp}, \& {Yung}}]{line2013b}
{Line}, M.~R., {Wolf}, A.~S., {Zhang}, X., {et~al.} 2013, \apj, 775, 137

\bibitem[{{Line} {et~al.}(2017){Line}, {Marley}, {Liu}, {Burningham}, {Morley},
  {Hinkel}, {Teske}, {Fortney}, {Freedman}, \& {Lupu}}]{line2017}
{Line}, M.~R., {Marley}, M.~S., {Liu}, M.~C., {et~al.} 2017, \apj, 848, 83

\bibitem[{{Lodders}(1999)}]{lodders1999}
{Lodders}, K. 1999, \apj, 519, 793

\bibitem[{{Lodders} \& {Fegley}(2002)}]{loddersfegley2002}
{Lodders}, K., \& {Fegley}, B. 2002, \icarus, 155, 393

\bibitem[{{Lueber} {et~al.}(2022){Lueber}, {Kitzmann}, {Bowler}, {Burgasser},
  \& {Heng}}]{lueber2022}
{Lueber}, A., {Kitzmann}, D., {Bowler}, B.~P., {Burgasser}, A.~J., \& {Heng},
  K. 2022, \apj, 930, 136

\bibitem[{{Madhusudhan} {et~al.}(2014){Madhusudhan}, {Amin}, \&
  {Kennedy}}]{Madhusudhan2014ApJ...794L..12M}
{Madhusudhan}, N., {Amin}, M.~A., \& {Kennedy}, G.~M. 2014, \apjl, 794, L12

\bibitem[{{Madhusudhan} {et~al.}(2017){Madhusudhan}, {Bitsch}, {Johansen}, \&
  {Eriksson}}]{Madhusudhan2017MNRAS.469.4102M}
{Madhusudhan}, N., {Bitsch}, B., {Johansen}, A., \& {Eriksson}, L. 2017,
  \mnras, 469, 4102

\bibitem[{{Madhusudhan} \& {Seager}(2009)}]{madhusudhanseager2009}
{Madhusudhan}, N., \& {Seager}, S. 2009, \apj, 707, 24

\bibitem[{{Marley} \& {Robinson}(2015)}]{marleyrobinson2015}
{Marley}, M.~S., \& {Robinson}, T.~D. 2015, \araa, 53, 279

\bibitem[{{Marley} {et~al.}(1996){Marley}, {Saumon}, {Guillot}, {Freedman},
  {Hubbard}, {Burrows}, \& {Lunine}}]{marley1996}
{Marley}, M.~S., {Saumon}, D., {Guillot}, T., {et~al.} 1996, Science, 272, 1919

\bibitem[{{Marley} {et~al.}(2021){Marley}, {Saumon}, {Visscher}, {Lupu},
  {Freedman}, {Morley}, {Fortney}, {Seay}, {Smith}, {Teal}, \&
  {Wang}}]{marley2021ApJ...920...85M}
{Marley}, M.~S., {Saumon}, D., {Visscher}, C., {et~al.} 2021, \apj, 920, 85

\bibitem[{{McKemmish} {et~al.}(2016){McKemmish}, {Yurchenko}, \&
  {Tennyson}}]{mckemmish2016}
{McKemmish}, L.~K., {Yurchenko}, S.~N., \& {Tennyson}, J. 2016, \mnras, 463,
  771

\bibitem[{{Molli{\`e}re} {et~al.}(2020){Molli{\`e}re}, {Stolker}, {Lacour},
  {Otten}, {Shangguan}, {Charnay}, {Molyarova}, {Nowak}, {Henning}, {Marleau},
  {Semenov}, {van Dishoeck}, {Eisenhauer}, {Garcia}, {Garcia Lopez}, {Girard},
  {Greenbaum}, {Hinkley}, {Kervella}, {Kreidberg}, {Maire}, {Nasedkin},
  {Pueyo}, {Snellen}, {Vigan}, {Wang}, {de Zeeuw}, \& {Zurlo}}]{molliere2020}
{Molli{\`e}re}, P., {Stolker}, T., {Lacour}, S., {et~al.} 2020, \aap, 640, A131

\bibitem[{{Mukherjee} {et~al.}(2022){Mukherjee}, {Fortney}, {Batalha},
  {Karalidi}, {Marley}, {Visscher}, {Miles}, \& {Skemer}}]{mukherjee2022}
{Mukherjee}, S., {Fortney}, J.~J., {Batalha}, N.~E., {et~al.} 2022, arXiv
  e-prints, arXiv:2208.14317

\bibitem[{{{\"O}berg} {et~al.}(2011){{\"O}berg}, {Murray-Clay}, \&
  {Bergin}}]{oberg2011}
{{\"O}berg}, K.~I., {Murray-Clay}, R., \& {Bergin}, E.~A. 2011, \apjl, 743, L16

\bibitem[{{Parmentier} \& {Guillot}(2014)}]{parmentierguillot2014}
{Parmentier}, V., \& {Guillot}, T. 2014, \aap, 562, A133

\bibitem[{{Phillips} {et~al.}(2020){Phillips}, {Tremblin}, {Baraffe},
  {Chabrier}, {Allard}, {Spiegelman}, {Goyal}, {Drummond}, \&
  {H{\'e}brard}}]{phillips2020}
{Phillips}, M.~W., {Tremblin}, P., {Baraffe}, I., {et~al.} 2020, \aap, 637, A38

\bibitem[{{Robinson} \& {Catling}(2012)}]{robinsoncatling2012}
{Robinson}, T.~D., \& {Catling}, D.~C. 2012, \apj, 757, 104

\bibitem[{{Rothman} {et~al.}(2010){Rothman}, {Gordon}, {Barber}, {Dothe},
  {Gamache}, {Goldman}, {Perevalov}, {Tashkun}, \& {Tennyson}}]{rothman2010}
{Rothman}, L.~S., {Gordon}, I.~E., {Barber}, R.~J., {et~al.} 2010, \jqsrt, 111,
  2139

\bibitem[{{Saumon} \& {Marley}(2008)}]{saumonmarley2008}
{Saumon}, D., \& {Marley}, M.~S. 2008, \apj, 689, 1327

\bibitem[{{Schwenke}(1998)}]{Schwenke1998}
{Schwenke}, D.~W. 1998, Faraday Discussions, 109, 321

\bibitem[{{Sousa-Silva} {et~al.}(2015){Sousa-Silva}, {Al-Refaie}, {Tennyson},
  \& {Yurchenko}}]{Sousa-Silva2015}
{Sousa-Silva}, C., {Al-Refaie}, A.~F., {Tennyson}, J., \& {Yurchenko}, S.~N.
  2015, \mnras, 446, 2337

\bibitem[{{Stock} {et~al.}(2018){Stock}, {Kitzmann}, {Patzer}, \&
  {Sedlmayr}}]{stock2018fastchem}
{Stock}, J.~W., {Kitzmann}, D., {Patzer}, A. B.~C., \& {Sedlmayr}, E. 2018,
  \mnras, 479, 865

\bibitem[{{Tennyson} \& {Yurchenko}(2018)}]{Tennyson2018}
{Tennyson}, J., \& {Yurchenko}, S. 2018, Atoms, 6, 26

\bibitem[{{Tennyson} \& {Yurchenko}(2012)}]{Tennyson2012}
{Tennyson}, J., \& {Yurchenko}, S.~N. 2012, \mnras, 425, 21

\bibitem[{{Tremblin} {et~al.}(2016){Tremblin}, {Amundsen}, {Chabrier},
  {Baraffe}, {Drummond}, {Hinkley}, {Mourier}, \& {Venot}}]{tremblin2016}
{Tremblin}, P., {Amundsen}, D.~S., {Chabrier}, G., {et~al.} 2016, \apjl, 817,
  L19

\bibitem[{{Weck} {et~al.}(2003){Weck}, {Schweitzer}, {Stancil}, {Hauschildt},
  \& {Kirby}}]{weck2003}
{Weck}, P.~F., {Schweitzer}, A., {Stancil}, P.~C., {Hauschildt}, P.~H., \&
  {Kirby}, K. 2003, \apj, 582, 1059

\bibitem[{{Yurchenko} {et~al.}(2011){Yurchenko}, {Barber}, \&
  {Tennyson}}]{Yurchenko2011}
{Yurchenko}, S.~N., {Barber}, R.~J., \& {Tennyson}, J. 2011, \mnras, 413, 1828

\bibitem[{{Yurchenko} \& {Tennyson}(2014)}]{Yurchenko2014}
{Yurchenko}, S.~N., \& {Tennyson}, J. 2014, \mnras, 440, 1649

\bibitem[{{Yurchenko} {et~al.}(2013){Yurchenko}, {Tennyson}, {Barber}, \&
  {Thiel}}]{Yurchenko2013}
{Yurchenko}, S.~N., {Tennyson}, J., {Barber}, R.~J., \& {Thiel}, W. 2013,
  Journal of Molecular Spectroscopy, 291, 69

\bibitem[{{Zalesky} {et~al.}(2019){Zalesky}, {Line}, {Schneider}, \&
  {Patience}}]{zalesky2019}
{Zalesky}, J.~A., {Line}, M.~R., {Schneider}, A.~C., \& {Patience}, J. 2019,
  \apj, 877, 24

\bibitem[{{Zalesky} {et~al.}(2022){Zalesky}, {Saboi}, {Line}, {Zhang},
  {Schneider}, {Liu}, {Best}, \& {Marley}}]{zalesky2022}
{Zalesky}, J.~A., {Saboi}, K., {Line}, M.~R., {et~al.} 2022, arXiv e-prints,
  arXiv:2206.01199

\bibitem[{{Zhang} {et~al.}(2021){Zhang}, {Snellen}, {Bohn}, {Molli{\`e}re},
  {Ginski}, {Hoeijmakers}, {Kenworthy}, {Mamajek}, {Meshkat}, {Reggiani}, \&
  {Snik}}]{zhangy2021_isotopejupiter}
{Zhang}, Y., {Snellen}, I. A.~G., {Bohn}, A.~J., {et~al.} 2021, \nat, 595, 370

\end{thebibliography}

\end{document}